\documentclass[%
amsmath,amssymb, aps,prfluids,
]{revtex4-2}

\usepackage{graphicx}
\usepackage{dcolumn}
\usepackage{bm}
\usepackage{xcolor}

\begin{document}

\title{From deep to shallow water 2D wave turbulence: Emergence of soliton gas}
\author{Thibault Leduque, Maxime Kaczmarek, Herv\'e Michallet, Eric Barth\'elemy, Nicolas Mordant}
\email{nicolas.mordant@univ-grenoble-alpes.fr}
\affiliation{Laboratoire des Ecoulements G\'eophysiques et Industriels, Universit\'e Grenoble Alpes, CNRS, Grenoble-INP,  F-38000 Grenoble, France}%

\date{\today}
\begin{abstract}
Experiments on 2D random water wave
propagation in a large wave tank are analyzed when the effect of dispersion changes. A stereoscopic profilometry technique is used to measure the water surface displacement resolved in both time and space over a significant fraction of the wave tank. The wave regimes are characterized by analyzing the space-time spectral statistical properties of the wave field. At a given, finite, water depth, the effect of dispersion can be varied by tuning the peak frequency of the wave generation. In shallow water conditions, the waves are only weakly dispersive and this enables the propagation of solitons. {In these conditions random wave forcing produces soliton gases}. In deep water conditions, the waves are dispersive and for wideband spectra, one observes the development of weak turbulence. A transition between these regimes is observed when changing the peak forcing frequency (dispersion) and the wave amplitude (nonlinearity), with a clear threshold between states with solitons and soliton-less states. The development of a soliton gas is associated with a strong change of the wave spectrum as well as a significant evolution of the distribution of the water elevation. We also observed a strong effect of the finite size of the tank due to the peculiar reflection laws of line solitons.
\end{abstract}

\maketitle

\section{Introduction}

Solitons are nonlinear waves that have fascinated scientists since first being observed by John Scott Russell \cite{scott1844report}, due to their quasi-particle character. Their properties are fundamentally related to integrability through the Inverse Scattering Transform theory (IST) that enables one to decompose any free surface wave profile into a spectrum of solitons and dispersive cnoidal waves \cite{Gardner,Dauxois}. The IST spectrum is invariant in time. For 1D propagation in shallow water, the Korteveg-de Vries equation is the archetypal nonlinear wave propagation equation with integrable dynamics \cite{Dauxois}, its deep water counterpart being the non linear Schrödinger equation that describes the dynamics of nonlinear modulation of a carrier wave~\cite{gelash_bound_2019}. Once the peculiar interaction (collision) properties of pairs of soliton have been characterized, it is tempting to investigate the statistical properties of a gas of solitons made of a larger number of such "particles". The random soliton gas has also been called ``integrable turbulence'' by Zakharov \cite{Zakharov} due to its random nature and to its dynamics being ruled by an integrable equation (that involves an infinite number of conserved quantities). 
As solitons behave to some extent as particles, whose collisions are well understood, it has led to the development of a kinetic theory of soliton gases. This was initiated by Zakharov \cite{zakharov1971kinetic} for dilute gases and has been generalized in the last 20 years to dense gases of various kinds \cite{el2005kinetic,el2011kinetic,GEL}. Over the same time span, several experimental soliton gases have been implemented for various water waves or in fiber optics (see \cite{Suret} for an extended review of theoretical and experimental studies). The unidirectional 1D solitons of the KdV equation have been extended to 1D bidirectional ones by the Kaup-Boussinesq model \cite{Zhang} and to 2D cases with weak directional spreading by the Kadomtsev-Petviashvili (KP) model \cite{KP,Kodama2010}, see also \cite{ablowitz1981solitons,biondini2007line,infeld2000nonlinear,kodama2018solitons}. In the KP model, single solitons are line solitons with a KdV-like profile and an infinite extension in the transverse direction. However, due to their angles of propagation, the interaction of pairs of solitons is more complex \cite{Miles77a,Miles77b,WangP,KodamaYeh}. The soliton gas theory has been extended to the KP model recently \cite{Bonnemain}. Let us recall that a single 1D KdV soliton has a free surface displacement profile in time ($t$) and space ($x$) given by, 
\begin{equation}
\eta(x,t) = a\, \textrm{sech}^2(\beta(x-c t+\phi))
\end{equation}
with $a$ the soliton amplitude, $1/\beta$ the width of the soliton, $c$ its speed and $\phi$ a phase. For a KdV soliton one has $\beta=\sqrt{{3a}/{4h_0^3}}$ and $c=c_0 \, (1+a/2 h_0)$ where $h_0$ is the water depth {at rest} and $c_0=\sqrt{g h_0}$. Note that the velocity of a soliton is larger than $c_0$, which is the maximum velocity of linear long waves, in the very shallow water limit. The two solitons solution has been described for instance in \cite{Zhang} and the asymptotic effect of soliton collisions is to change the phase $\phi$, so that the effect of cumulative collisions can alter the effective speed of the soliton, and this effective speed lies at the core of the kinetic theory (see \cite{GEL} for details).

Despite being called ``turbulence'', integrable turbulence is intrinsically different from the more classical Navier-Stokes turbulence or the somewhat more related weak turbulence. In a soliton gas, due to the infinite number of conservation laws, no energy cascade can develop for instance. Weak turbulence concerns also nonlinear waves and is actually often referred to as ``wave turbulence'' (although wave turbulence can be defined as a wider framework) \cite{nazarenko2011wave,newell_wave_2011}.  The weak turbulence theory (WTT) describes mainly an out of equilibrium turbulent state in which energy cascades to small scales where it is dissipated into heat. In the WTT, energy is exchanged between sine waves through wave resonances as originally described by Hasselmann \cite{Hasselmann}. For gravity surface waves, the resonances involve four waves that fulfill the resonance conditions 
\begin{equation}
\mathbf k_1+\mathbf k_2=\mathbf k_3+\mathbf k_4 \quad\textrm{and} \quad \omega_1+\omega_2=\omega_3+\omega_4 
\end{equation}
with the wave vectors $\mathbf k_i$ and the frequencies $\omega_i$ being related by the dispersion relation. In deep water, in the limit of infinite size of the system and for weak nonlinearity, the addition of the exchanges through numerous resonances results in a direct energy cascade and an inverse wave action cascade. Under ideal conditions, stationary spectra have been predicted (Kolmogorov-Zakharov spectra \cite{nazarenko2011wave}). {In practice, however, the outcome is typically a broadband spectrum, which is often represented using the JONSWAP parametrization} \cite{jonswap}. Note that such resonances can only occur for 2D propagation since the 4-wave nonlinear coupling coefficient vanishes for collinear waves \cite{Zak1D}. In 1D, a higher order 5-wave resonance may replace it \cite{Zak1D,zakharov1980degenerative,zakharov1988additional} but, since the nonlinear time scale of turbulence development is much longer, its observation in laboratory conditions is impeded by dissipation. For 1D propagation of narrow-banded spectra, another nonlinear phenomenon can operate: the modulational or Benjamin-Feir instability involving near-resonant waves, that leads to development of envelop solitons. However, it has been observed that this instability does not operate when the directivity is wide and when the depth is too low~\cite{janssen07,onorato09,zve} so that this phenomenon is not relevant in our conditions. 

In practice, the wave dynamics are not strictly integrable due to viscous energy dissipation. This questions the relevance of integrable theory for real waves. A very weak dissipation (compared to the typical duration of either soliton collision or the transient of evolution of the soliton gas) may allow the temporal evolution to be ruled by integrability at intermediate time scale. We observed such dynamics in 1D soliton gas in shallow water \cite{RedorPRL,RedorEIF,RedorPRF}. However weak dissipation and non linearity are also a prerequisite for the development of weak turbulence. A key feature to discriminate between 
{weak turbulence and integrable turbulence}
is dispersion. Indeed, single soliton propagation requires a balance between weak dispersion and weak nonlinearity to maintain the shape of the soliton unchanged during the propagation. For water surface gravity waves, the dispersion is set by the water depth. The linear dispersion relation is:
\begin{equation}
\omega^2 = g \, k \tanh (k h_0)             \label{eq-rd}
\end{equation}
where $\omega$ is the angular frequency, $g$ the gravity acceleration, $k$ the wave-number and $h_0$ the water depth at rest. Frequency dispersion is ruled by the non-dimensional parameter $kh_0$. For deep water ($kh_0\gg 1)$) surface waves are (strongly) dispersive since (\ref{eq-rd}) reduces to $\omega^2=g \, k$ and for very shallow water ($kh_0\ll 1)$) the waves are non dispersive with $\omega^2 =c_0^2 \, k^2$. As the dispersion parameter $kh_0$ reduces, the waves change gradually from dispersive to weakly dispersive. Further complexity arises for wide band wave spectra as part of the spectrum can be considered in deep water while low frequencies may correspond to a shallow water regime. This situation arises in coastal areas where the water depth changes progressively when nearing the shore. In that intermediate case, integrable dynamics and weak turbulence could operate simultaneously depending on length and time scales.

Non-linearity is also affected by the depth. For illustration, consider a 1D monochromatic Stokes wave expansion truncated at second order in wave amplitude following \cite{stokes1847,whitham74,dean1991,toffoli2007}:
\begin{equation}
\eta(x,t) =  a \cos(kx-\omega t)+ \frac{1}{4} a^2 K^+\cos(2(kx-\omega t))\, ,
\end{equation}
where $a$ is the amplitude of the fundamental wave and with the coupling coefficients:
\begin{equation}
K^+ = k \, \frac{\cosh{(k h_0)}}{\sinh^3{(k h_0)}} \, \left( 1 + 2\cosh^2{(k h_0)} \right) 
\end{equation}\, 
The expression $a K^+/4$ is the dimensionless ratio of the amplitude of the second harmonics to that of the initial wave. Note that the second harmonic does not comply to the linear dispersion relation (\ref{eq-rd}); it is a bound wave as discussed in Appendix \ref{sec.bound} in a more general case. For deep water ($kh_0 \gg 1$), one gets $aK^+/4 \approx k\, a/2$ 
which is proportional to the steepness $k a$ of the monochromatic wave. The steepness of the wave is thus the parameter of nonlinearity in deep water. In shallow water ($kh_0 \ll 1$), one gets $\frac{aK^+}{4} \approx  \frac{3 k a}{4(kh_0)^3}$ which is the Ursell number $Ur$. The Ursell number is defined as the ratio of the nonlinearity $a/h_0$ over the the dispersion $(kh_0)^2$ in the context of shallow water equations \cite{ursell1953}. At a given (small) value of $k a$ the Ursell number diverges when $kh_0$ decreases. The nonlinear effects are thus strongly increased for shallow water, to the extent that the truncation at second order is not relevant as higher orders need to be taken into account as well.
Indeed one expects solitons or cnoidal waves to emerge when $Ur$ is of order one. Zhao {\it et al.} \cite{zhao2024} define the limit between regimes of cnoidal waves and regimes of possible Stokes expansion as $Ur=0.25$ (using our definition of $Ur$, $Ur=26$ with their definition, which differs by a factor $3/(32\pi^2)$, see \cite{zhao2024} and references therein).
 
In summary, in deep water, the dynamics are dominated by dispersion so that one can expect to observe weak turbulence and no KdV-like solitons can propagate. In shallow water, dispersion weakens and nonlinearity is reinforced so that solitons can emerge. Our present work investigates the intermediate depth regime where the Ursell number is close to the limit $Ur=0.25$ either above or below this critical value. The competition between weak turbulence and soliton gas was studied using a very wide wave tank that enables a 2D propagation of the waves. The 2D character is of major importance to make sure that weak turbulence can develop at large depth. We have developed a time and space resolved measurement of the water surface~\cite{leduqueEIF} to describe the evolution of the statistical properties of randomly forced waves. The dispersion is varied by changing the frequency of the forcing at constant depth and the nonlinearity is modified by tuning the magnitude of the forcing. {We address the question of the impact of the emergence of solitons on the statistical properties of wave turbulence, specifically on the power spectrum density and the distribution of the water surface displacement.}

In the following sections, we first describe the facility, the measurements and the wave forcing. In the next part, we describe the statistical properties when dispersion and nonlinearity are varied. Finally we discuss the impact of the finite size of the wave tank on the dynamics of the waves. 

\section{Experimental setup}

\subsection{Wave facility}

Our research team has access the multi-directional wave tank LHF facility of the ARTELIA company, located in Pont de Claix, France. The size of the tank is 27~m by 30~m. For all experiments, the water depth is $h_0=35\pm 0.4$~cm (within a few millimeters due to bottom irregularities). One 30~m side (corresponding to $x=0$) is equipped with 60 piston-type wavemakers (50~cm wide) that can be driven independently so as to generate arbitrary waves. The maximum stroke of the wavemakers is $\pm 30$~cm, which restricts the magnitude of the solitons that can be generated. The tank has vertical walls on all other sides so that boundary conditions are reflective. This was implemented so that waves can bounce back and forth to allow for sustained nonlinear interactions with other waves.

\begin{figure*}[!htb]
(a) \includegraphics[width=0.8\textwidth]{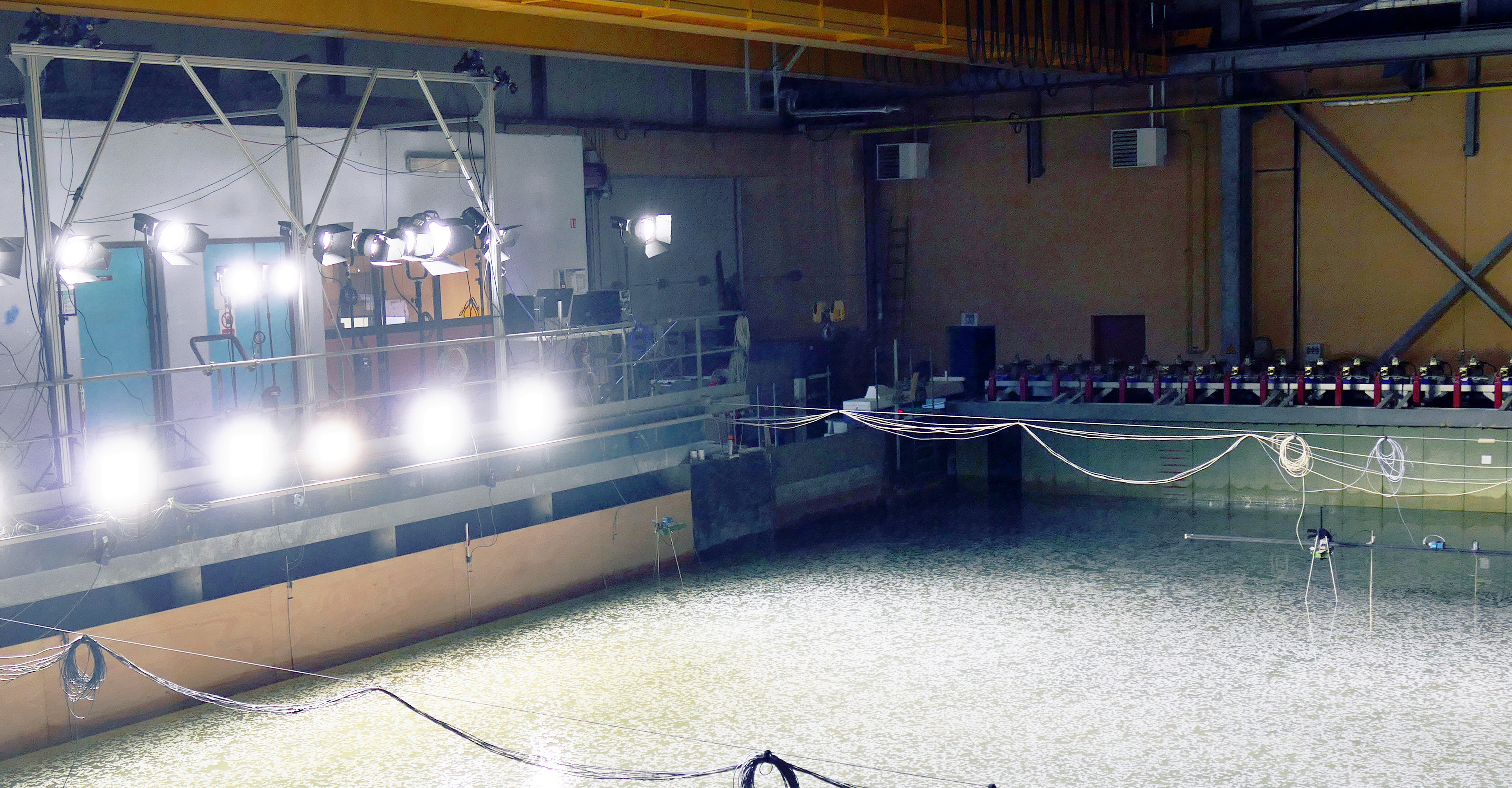}

(b) \includegraphics[height=8cm]{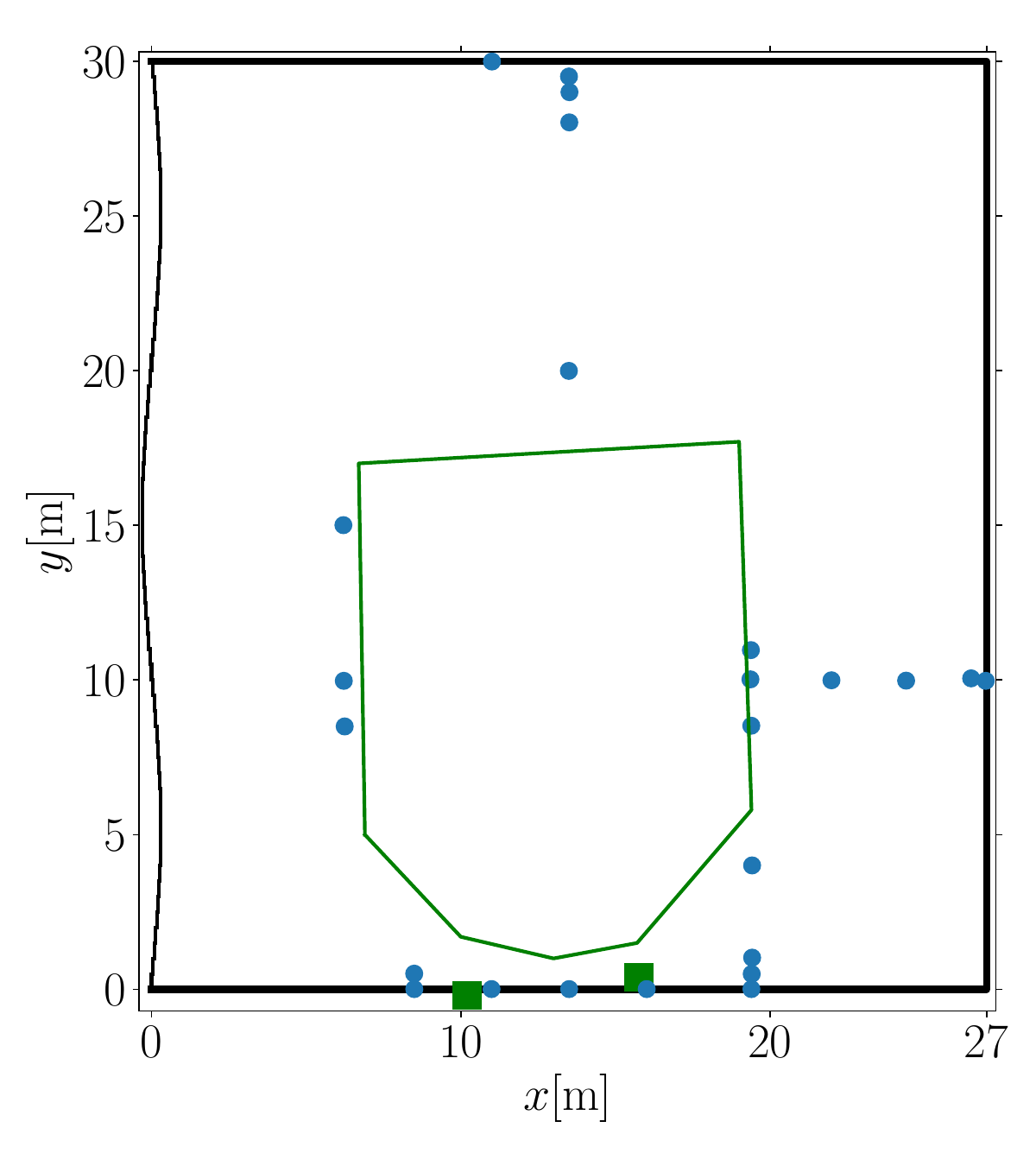}(c) \includegraphics[height=8cm]{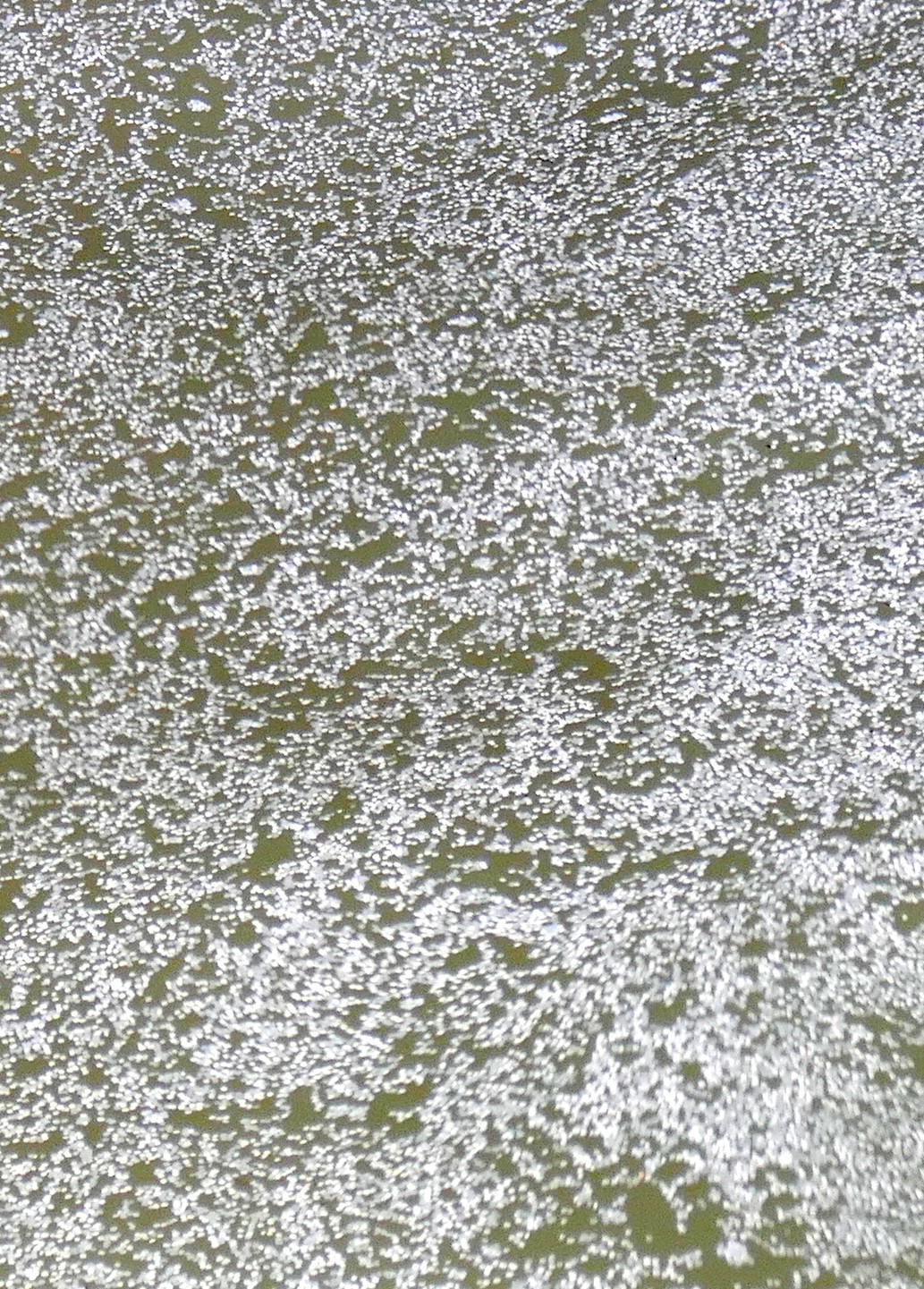}
\caption{\label{probes} (a) View of part of the ARTELIA LHF wave tank with the camera systems. The  pairs of cameras can be seen at the top of the aluminum frame at the upper-left part of the image (another pair of cameras were installed but not used in this paper). The water surface appears white due to the floating particles. A few wavemakers can be seen on the right of the image. The cables carry the signal of the capacitive wave probes. A pair of such probes is visible at the bottom right of the image. The field of view of the cameras lies in between the two lines of cables separated by $12\,$meters. A set of spotlights illuminate the water surface at small incidence angles so as to avoid specular reflections into the cameras.
(b) Top view sketch of the tank. Blue dots: positions of the capacitive wave probes. The green polygon delineates the measurement region of the stereoscopic system. The green squares are the positions of the cameras. The wavemakers are on the left at $x=0$. (c) View of a region of the water surface with the floating particles.}
\end{figure*}

\subsection{Wave measurements} 
The wave measurements have been described in \cite{leduqueEIF}. Here, the focus is thus on the main characteristics and modifications compared to \cite{leduqueEIF} that correspond to a preliminary experimental campaign. Wave fields are measured through two systems. First, a set of 24 capacitive wave gauges provide single point accurate measurements of the water surface elevation sampled at $64\,$Hz with an accuracy of about 0.5~mm. Second, a stereoscopic measurement system is used with a pair of cameras that provide time and space resolved measurements of the water surface elevation over a sub-region of the wave tank. The cameras (JAI SP-12000-CXP4, 12 MPixels) are tuned to view an area equal to about 180~m$^2$ from about 2 meters from the wall up to the center of the tank. A sketch of the measurements is shown in Fig.~\ref{probes}. The cameras are positioned along the wall at $y=0$, about 5.5~m above the water surface, and separated by 5.5~m in the $x$ direction (see details in \cite{leduquePhD2024}). As described in \cite{leduqueEIF}, the water surface is made visible for the camera by seeding a large concentration of buoyant plastic particles of size roughly 3~mm. This generates texture at the water surface that can be used by an image correlation technique to reconstruct the water surface elevation. The presence of the particles is observed to induce a slight damping of the waves at frequencies higher than 1~Hz (see Fig.4.b of \cite{leduqueEIF}) but these effects remain weak and may also be due to surfactants released by the particles rather than to the particles themselves \cite{leduqueEIF,Campagne}. 
{In any case, the presence of particles does not affect the dispersion relation at the considered frequencies (below that of capillary waves).}
The frame rate of the cameras is $10\,$fps except for the experiments with a peak frequency of the forcing of $1\,$Hz for which the frame rate is $20\,$fps. The vertical resolution of the video stereoscopic system is a couple of millimeters and the minimum wavelength that can be measured with our image processing parameters is 34~cm.

The stereoscopic systems provides a measurement of the wave induced free surface displacement $\eta(x,y,t)$. Various statistical quantities can be deduced such as the water surface displacement power spectrum density in the frequency space $E(\omega)$ or in wave vector-frequency space $E(k_x,k_y,\omega)$. To compute $E(k_x,k_y,\omega)$, Fourier transforms in space are computed over the whole measurement domain (over which the image processing quality is good enough) and a Hanning window in both $x$ and $y$ directions adapted to the shape of the spatial domain.
{Averaging is performed along the time direction assuming ergodicity, as is routinely done in the analysis of statistically stationary turbulent signals. This assumption seems reasonable because the duration of the records is much longer than the nonlinear evolution of the waves, and the statistical convergence is checked to be similar for all points in the video-recorded area.} The classical Welch method is used to compute Fourier transforms in time using Hanning windows as well.
$E(\omega)$ is obtained by summing over the wave vector dimensions. Water surface displacements probability density functions (PDF) $P(\eta)$ are evaluated using histograms and by averaging over space. Equivalent $E(\omega)$ and $P(\eta)$ can be computed from the capacitive probes by averaging over the ensemble of probes. The comparison of $E(\omega)$ between capacitive probes and stereoscopic system shows a good agreement up to frequencies about $1.5\,$Hz (see Appendix \ref{sec.compsp}). In the following sections, we choose to show the $E(\omega)$ spectra obtained from the capacitive probes due to a better accuracy at high frequency. However, the probability distributions $P(\eta)$ are computed from the stereoscopic system because of the much higher amount of data allowed by the spatial measurement as compared to the local probes.

\subsection{Wave forcing}

\begin{figure}[!htb]
\includegraphics[width=10cm]{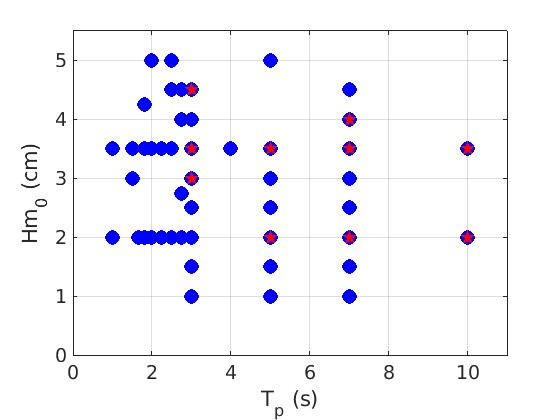}
\caption{\label{fig:params} Parameters $H_{m0}$ and $T_p$ of the JONSWAP spectra forcings of all experiments (see text for definitions of the parameters). Symbols correspond to different directivity of the forcing: blue disks to $s_{max}=30$ and red stars to $s_{max}=1$.}.
\end{figure}

{During the experiments reported here, the waves were continuously generated
using a random forcing complying a JONSWAP spectrum.} This spectrum is a parametrization of natural waves measured in the North Sea as reported in \cite{jonswap}. The details of the analytic expression for the JONSWAP spectrum are given in appendix \ref{app:jonswap}. The spectrum is made of a frequency content with a relatively narrow band spectrum peaked at a $f_p$ that was varied between 0.1~Hz and 1~Hz (peak periods $T_p=1/f_p$ between 1~s and 10~s). At frequencies higher than $f_p$ the JONSWAP forcing decays as $1/f^5$ (up to 2 Hz at which the spectrum is truncated). The spectrum contains also a directional component characterized by a spreading factor $s_{max}$ (see appendix~\ref{app:jonswap}). For most of the experiments reported here $s_{max}=30$ which provides a relatively narrow directional wave beam. The angular width at $f_p$ has a width at half height equal to $\pm 17^\circ$. A few experiments were undertaken with $s_{max}=1$ which corresponds to a width $\pm 90^\circ$ at $f_p$, so a forcing close to isotropic. Henceforth, when not specified, the value of $s_{max}$ is 30. The last main parameter is the magnitude of the forcing that is usually characterized by the significant wave height $H_{m0}$ which equals 4 times the water surface displacement standard deviation $\eta_{rms}$. This would be the water surface displacement that would be measured if the waves were emitted in open waters or with absorbing beaches. However, due to reflections on the walls, the actual water surface displacement is not $H_{m0}/4$ but closer to $0.7 H_{m0}$ (see below). The values of the parameters of the experiments $H_{m0}$ and $T_p$ are shown in Fig.~\ref{fig:params}. The JONSWAP spectrum is relevant for deep water waves but we use it here as a conveniently parameterized random forcing which is relatively narrow banded (the width at half height of the elevation spectrum is about 20\% of $f_p$) without being monochromatic. 

A typical experimental sequence lasts about 30 minutes with the following sequence: (i) the recording of signals of the capacitive wave gauges is first started, (ii) slightly later, at $t=0$ the wavemakers motion is initiated, (iii) {at $t=10$~minutes when the wave field is deemed statistically stationary the recording of the camera images is started for 20 minutes.}
The wavemakers are stopped once the recording completed.
The frame rate of the cameras being 10~fps it corresponds to two times 12,000 images so 288~GB of raw 8-bit images. The total volume of raw data stored during the present experiments is close to 20~TB. 

By varying the $f_p$ frequency, one tunes the effective water depth by changing the parameter $k_p\,h_0$ where $k_p$ is the peak wavenumber computed with the linear dispersion relation(\ref{eq-rd}). By changing the peak frequency in the range $[0.1,1]$~Hz, $k_p$ varies in the range $[0.34,4.4]$~rad/m so that $k_ph_0$ lies in $[ 0.12, 1.5]$. At the lowest frequency ($f_p=0.1$~Hz) the waves are clearly in the shallow water regime while at the highest frequency ($f_p=1$~Hz) the waves are more in the deep water regime. Thus our experiments can scan the effect of the change of depth regime on the statistical properties of the waves.

{The forcing was kept significantly lower to what it would have been, had absorbing beaches been used instead of vertical walls. 
Wave amplitudes were restricted in order to avoid strong wave breaking. These restrictions prevent the appearance of large areas without particles that occur for large breaking events and render the stereoscopic measurement impossible.
Because of the continuous forcing and the multiple wave reflections, scarce and weak
white-capping events were however observed in the video-recorded area for the strongest forcing conditions.
Wave breaking often occurs close to the walls and in the corners of the tank where the wave amplitudes are often significantly larger than in the bulk of the tank, due to the peculiar reflection laws of the solitons that are discussed at the end of the article. Theoretically, in the worst case, soliton amplitudes can be amplified by a factor 4 near the wall.}


\section{Effect of the dispersion}
\begin{figure}[!htb]
\includegraphics[width=\textwidth]{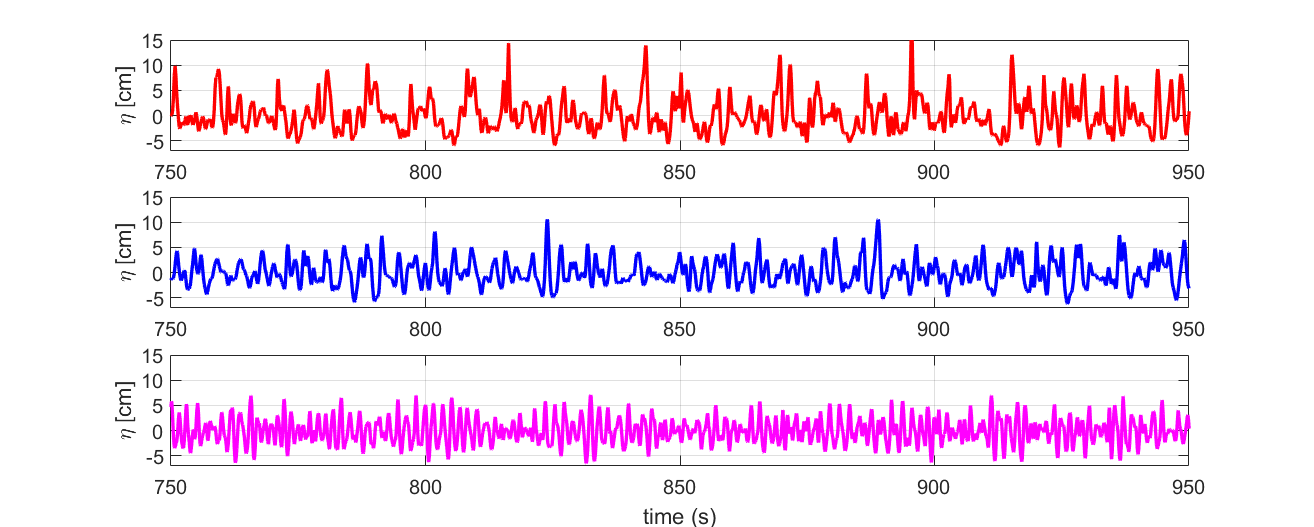}
\caption{\label{fig:eta35} Examples of water surface displacement time series for the experiments at $T_p=1.8$, $4$ and $10$~s from bottom to top as described in table~\ref{table_Tp}. The surface displacement has been measured by a capacitive probe located at $x=19.37$~m and $y=10.02$~m. }
\end{figure}

\begin{figure}[!htb]
\includegraphics[width=\textwidth]{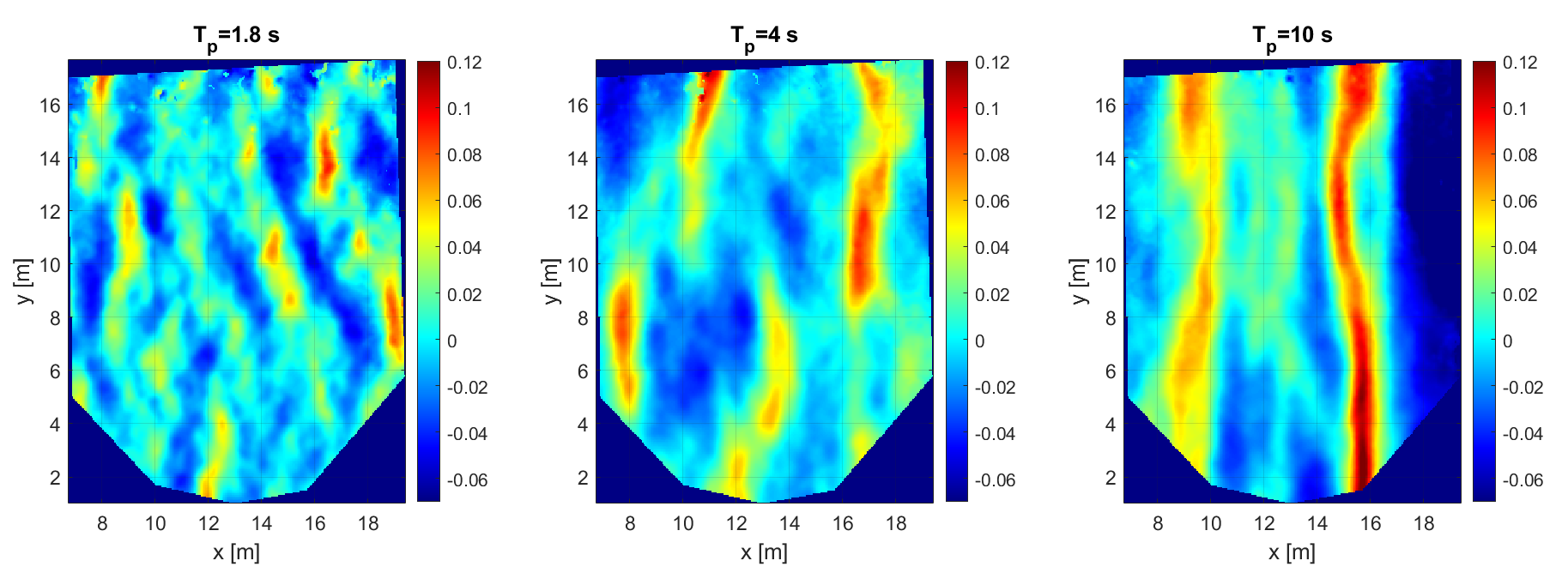}
\caption{\label{fig:eta35stereo} Examples of snapshots of the wave fields for the experiments at $T_p=1.8$, $4$ and $10$~s from left to right as described in table~\ref{table_Tp}. The water surface displacement is measured by the stereoscopic reconstruction system. The water surface displacement is coded in linear color scale in meters.}
\end{figure}

In a first series of experiments, we wanted to investigate the effect of varying the dispersion factor $k_ph_0$ while keeping the amplitude of the forced wave $H_{m0}/h_0$ constant. A set of experiments is listed in table~\ref{table_Tp} for $H_{m0}=3.5$~cm, which corresponds to $H_{m0}/h_0=0.1$. Examples of time series of the water surface displacement measured by one of the capacitive probes are shown in Fig.~\ref{fig:eta35} and snapshots of the water surface displacement obtained by the stereoscopic system are shown in Fig.~\ref{fig:eta35stereo}. These three examples show very different wave field properties. At the lowest frequency forcing ($T_p=10$~s), waves are strongly peaked with large maximum values and they exhibit long crests with a spatial coherence that extend over a large fraction of the width of the tank. 

\begin{table*}[!htb]
\begin{ruledtabular}
{\begin{tabular}{lcccccccccccc}
$T_p$ (s)& 1 & 1.5 & 1.8 & 2 & 2.25 & 2.5 & 3 & 4 & 5 & 7 & 10\\
$\eta_{rms}$ (cm) & 2.3 &2.4&2.4&2.4&2.5&2.4&2.5&2.5&2.7&2.8&2.7\\
$S$ skewness & 0.24 & 0.25 & 0.29 & 0.32 & 0.34 & 0.37 & 0.47 & 0.60 & 0.72 & 0.94 & 0.83\\
$K$ flatness & 3.1 & 3.1 & 3.1 & 3.1 & 3.2 & 3.3 & 3.3 & 3.6 & 4.0 & 4.6 & 4.4\\
$k_ph_0$ & 1.54 &0.88&0.71&0.63&0.55&0.49&0.41&0.30&0.24&0.17&0.12\\
$Ur=\dfrac{3\eta_{rms}}{4k_p^2h_0^3}$  & 0.02&0.07&0.10&0.13&0.17&0.21&0.32&0.58&1.00&2.1&4.1\\
$E_{solitons}/E_{free}$ & 8.5e-3 & 0.10 & 0.14 & 0.20 & 0.24 & 0.28 & 0.75 & 1.3 & 2.3 & 4.9 & 4.8\\ 
\end{tabular}}
\end{ruledtabular}
\caption{\label{table_Tp}Parameters of the experimental runs with constant significant wave height $H_{m0}=3.5$~cm measured by the stereoscopic system. $T_p$ is the peak period of the forcing spectrum, $\eta_{rms}$ is the measured root mean squared elevation, $S$ is the skewness of the water surface displacement, $K$ is the flatness factor, $k_ph_0$ is the dispersion parameter, $Ur$ is the Ursell number and $E_{solitons}/E_{free}$ is the ratio of energy on the soliton dispersion relation to that of the free wave dispersion relation at $f=1.5$~Hz. See text for more details.}
\end{table*}

\begin{figure}[!htb]
\includegraphics[width=8cm]{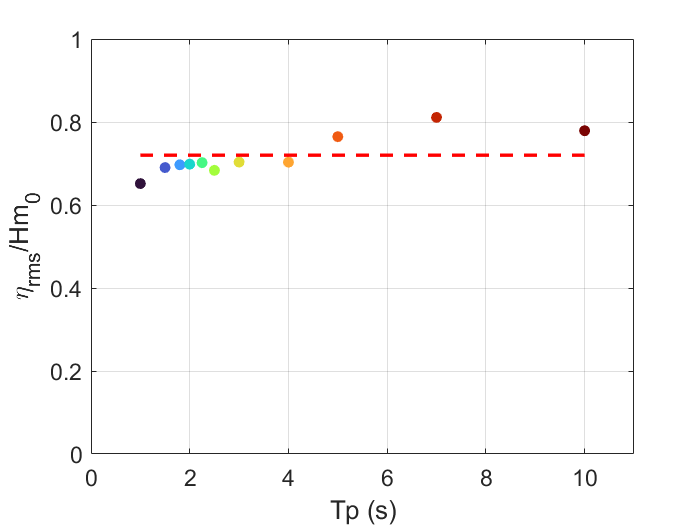}
\caption{\label{Hm35rms} Wave amplification $\eta_{rms}/H_{m0}$ for the experiments with $H_{m0}=3.5$~cm described in table \ref{table_Tp}. The red dashed horizontal line is $0.72$.}.
\end{figure}

The {\it rms} water surface displacement is shown in Fig.~\ref{Hm35rms}. Although the wave induced free surface displacement slightly increases with $T_p$ it is seen to remain very close to $\eta_{rms}=0.72 \, H_{m0}$ within 10\% variability so it can be considered as constant for this set of experiments. It means that $\eta_{rms}$ is about 3 times larger that it would be if the waves were generated in a tank with absorbing beaches.

\subsection{Spectra}
\begin{figure*}[!htb]
(a)\includegraphics[width=\textwidth]{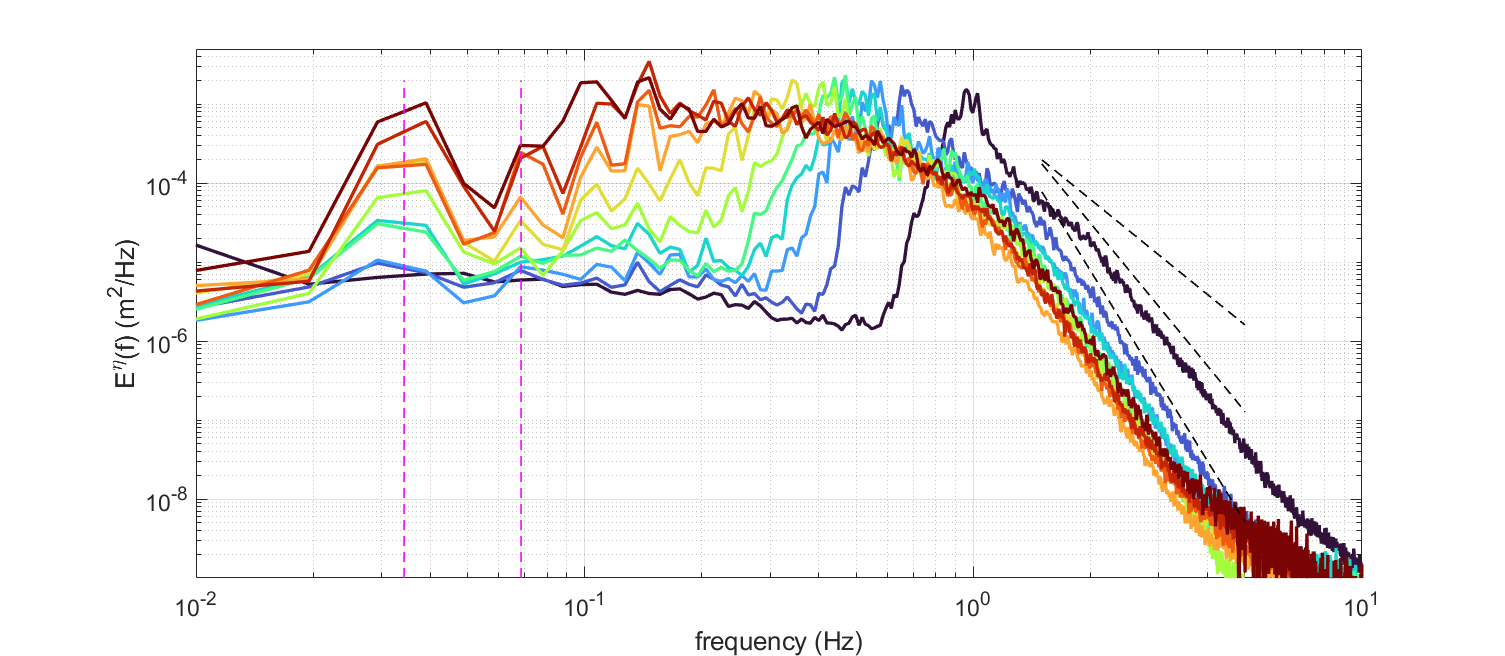}
(b)\includegraphics[width=\textwidth]{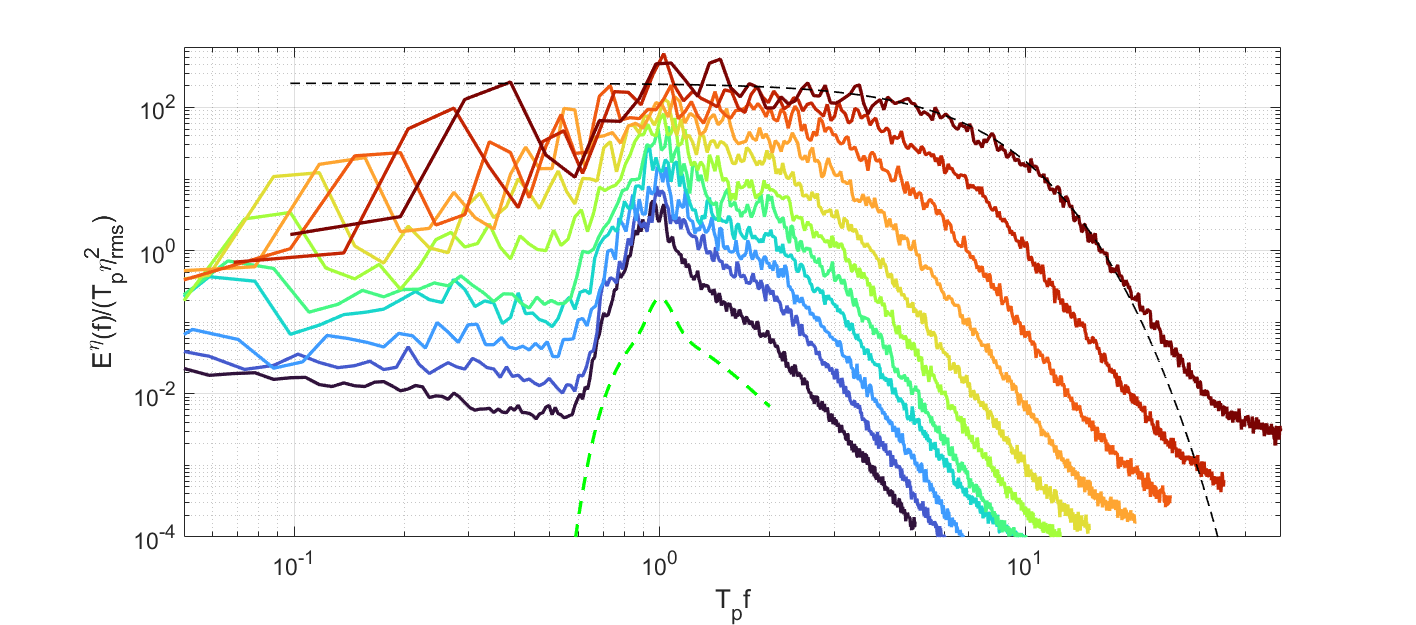}
\caption{\label{spf35}Frequency spectra for $H_{m0}=3.5$~cm as a function of $T_p$, estimated from the data of capacitive probes away from any wall by more than 5~m (8 probes). Colors from black to dark red/brown correspond to increasing value of $T_p$ from 1 to 10~s (see table~\ref{table_Tp}). (a) shows dimensional spectra and (b) shows dimensionless spectra. (a): the vertical magenta dashed lines indicate the frequency of the seiching modes with $(k_x,k_y)=(\pi/L_x,0)$ and of the mode $(k_x,k_y)=(2\pi/L_x,0)$, with $L_x=27$~m the length of the tank along $x$ direction. These modes are the first and second resonant modes in the $x$ direction, corresponding to a half and a single wavelength respectively along $x$ direction and none along $y$. The black dashed lines correspond to power law decays of $1/f^\alpha$ with $\alpha=4,6,8$. (b): the spectra are shifted vertically by factors of 2 for visual clarity. The black dashed line is a fit of the spectrum at $T_p=10$~s with an the approximation $E(f)\propto f^2/\sinh(\alpha f)^2$. The green dashed line is the spectrum of the forced waves at $T_p=1$~s shifted below the corresponding measured spectrum by a factor $2.7^2$ ($2.7$ corresponds to the ratio of the measured $\eta_{rms}$ over $H_{m0}/4$).}
\end{figure*}

Frequency spectra of the water surface displacement are shown in Fig.~\ref{spf35}, both in dimensional units and in dimensionless form, computed from the capacitive probes signals (spectra from the stereoscopic system are very similar, see appendix~\ref{sec.compsp}). For the experiments with $T_p=1$ or $1.5$~s, the spectra look similar to a JONSWAP spectrum. The decay of the spectrum for $f>f_p$ is compatible with a power law decay with an exponent between $6$ and $8$ (although the range is very narrow) which is typical of laboratory wave turbulence experiments in deep water with contaminated water surface (see \cite{Campagne,Deike}). This exponent is significantly higher than that of weak turbulence theory \cite{nazarenko2011wave} (that predicts an exponent equal to 4) or of field observations that show exponents close to the theoretical prediction (see for example \cite{Leckler,Lenain}). This has been interpreted as extra dissipation due to surface contamination by dust or surfactants which is very significant at frequencies higher than 1~Hz \cite{Campagne}. This dissipation is actually observed even without particles. The inflection at the highest frequencies ($f>2.5$~Hz) is most likely due to experimental noise. Note that the energy level is very weak at frequencies lower than the excitation. 

When $T_p$ is increased, the shape of the spectra changes significantly. In a first stage, an increasing peak or kink appears at $2f_p$. The low frequency part of the spectrum (for $f<f_p$) increases as well. This increase is visible down to $f_p\approx 0.034$~Hz which is the frequency of a very low frequency resonant mode of the tank with one half wavelength along the $x$ axis and none along the $y$ axis (a seiche mode along the $x$ axis). These changes are consistent with the second order effects that generate higher order harmonics as well as low frequencies (see Appendix \ref{sec.bound}). Decreasing the dispersion $k_ph_0$ reinforces the second order effects as discussed in the introduction. The snapshots of the water surface displacement, shown in Fig.~\ref{fig:eta35stereo}, confirm the observations from the spectra. The water surface displacement for $T_p=10$~s is made of long ridges mostly propagating along the $x$ direction and with a very long coherence in the $y$ direction that can be characterized as line solitons possibly described by the KP theory for instance. 
By contrast, the example of wave field at $T_p=1.8$~s has a much lower coherence in the $y$ direction (short crested waves) and is typical of anisotropic weak turbulence.

When $T_p$ is further increased, the Ursell number exceeds $0.25$ (experiments with orange to red curves with $T_p>3$~s) and the spectra evolve into a quite distinct shape. The spectra are very flat at low frequencies and follow a decay resembling $\propto f^2/\sinh(\alpha f)^2$ (Fig.~\ref{spf35}(b)) which is the shape of the squared Fourier transform of a single KdV soliton. At the largest $T_p=10$~s the forcing peak $fp$ is no more present, which mean that there is a disorganisation of the wave field with regards to the forcing. At large $T_p$, the spectra show a much larger bandwidth as compared to the JONSWAP spectrum. At $T_p=10$~s, the bandwidth is about 1 decade in frequency. This shape has strong resemblance with spectra observed in 1D random soliton gaz by Redor {\it et al.} \cite{RedorPRL,RedorPRF}. At large Ursell number in a linear wave flume, Redor {\it et al.} observed the formation of soliton gas with random positions and amplitudes. In these regimes, their spectra also display a flat spectrum followed by an exponential decay. The time series for $T_p=10$~s in Fig.~\ref{fig:eta35} shows that the signal contains sharp positive peaks and the signal is strongly asymmetric between peaks and troughs. Peaks are solitons that have been generated by a fission process of the low frequency wave forcing as identified by Zabuski \& Kruskal~\cite{ZK}, as observed by \cite{Trillo}.

\begin{figure*}[!htbp]
\includegraphics[width=\textwidth]{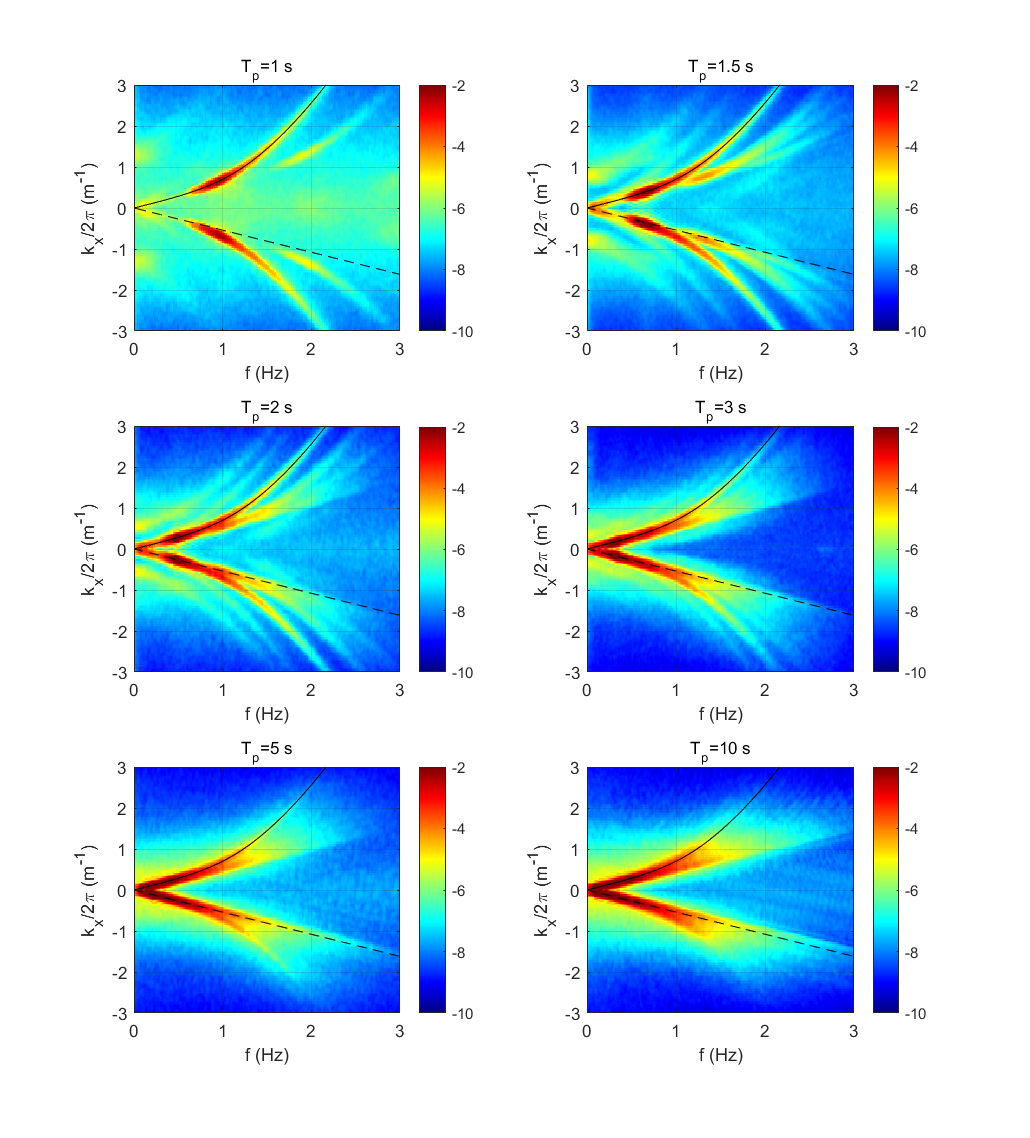}
\caption{\label{spkxf35} Spectra $E(k_x,0,\omega)$ for $H_{m0}=3.5$~cm and for various $T_p$. The dashed line is the linear shallow water dispersion relation $\omega^2=c_0^2k^2$ (the branch for positive $k_x$ is not shown for clarity). The solid line is the finite depth linear dispersion relation $\omega^2=gk\tanh(kh_0)$ (the branch for negative $k_x$ is not shown). The colorbar is $\log_{10}(E)$, $E$ in m$^4$s.}
\end{figure*}

\begin{figure*}[!htbp]
\includegraphics[width=\textwidth]{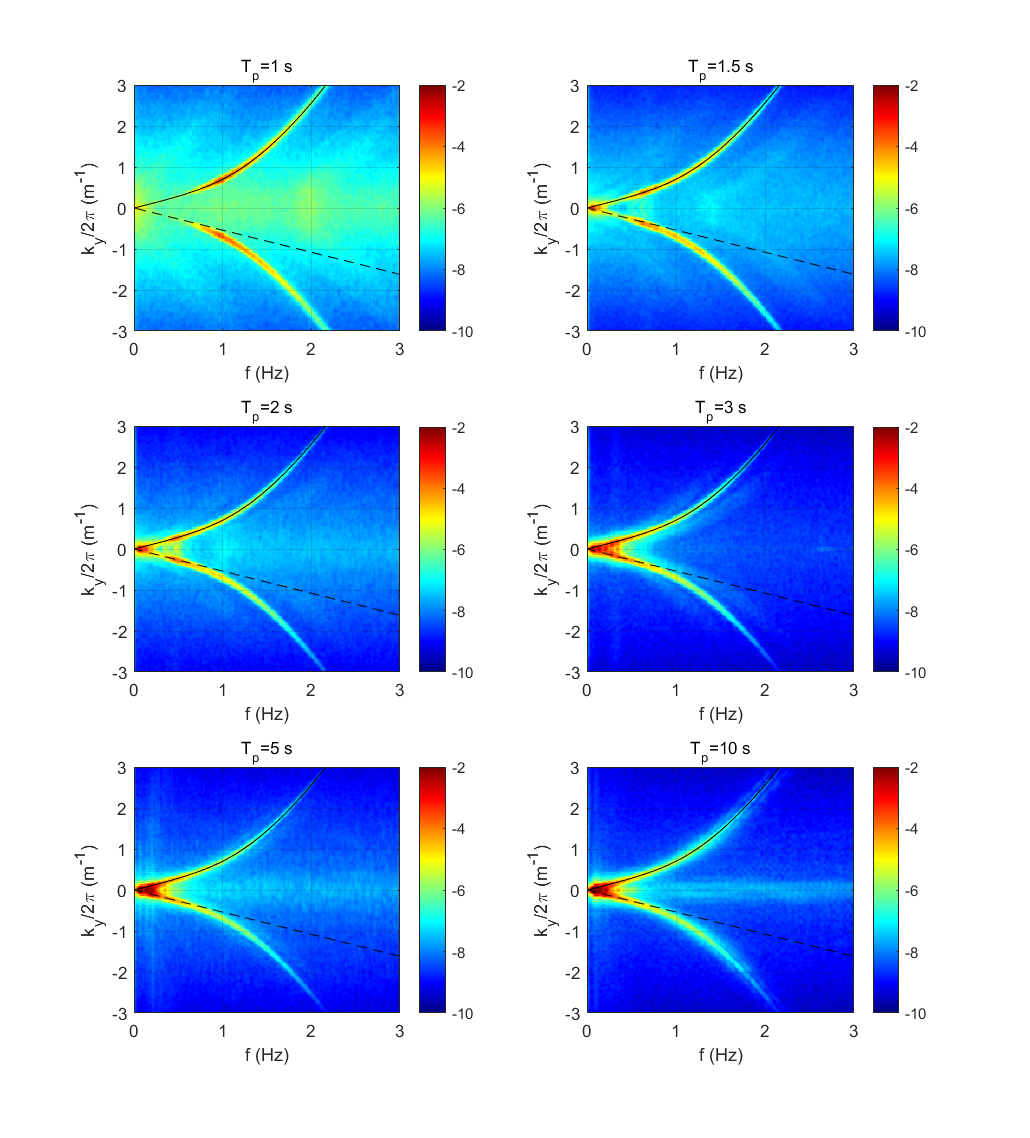}
\caption{\label{spkyf35} Spectra $E(0,k_y,\omega)$ for $H_{m0}=3.5$~cm and for various $T_p$. The dashed line is the linear shallow water dispersion relation $\omega^2=c_0^2k^2$ (the branch for positive $k_y$ is not shown for clarity). The solid line is the finite depth linear dispersion relation $\omega^2=g k \tanh(kh_0)$ (the branch for negative $k_y$ is not shown). The colorbar is $\log_{10}(E)$, $E$ in m$^4$s.}
\end{figure*}

\begin{figure*}[!htbp]
\includegraphics[width=\textwidth]{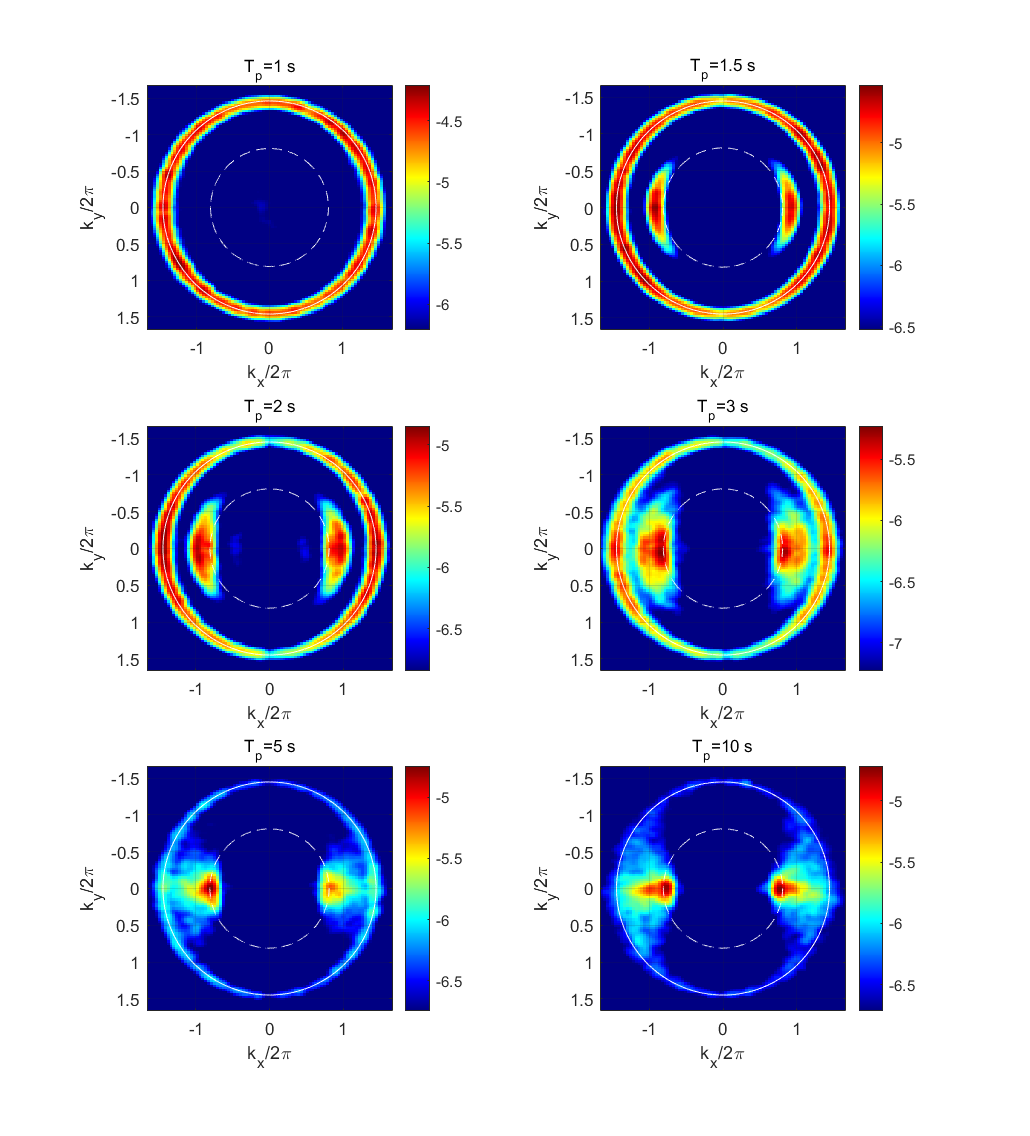}
\caption{\label{spkxky35} Spectra $E(k_x,k_y,\omega=3\pi)$ for $H_{m0}=3.5$~cm and for various $T_p$. The inner white dashed circle is the linear shallow water dispersion relation $\omega^2=c_0^2k^2$. The outer white circle is the finite depth linear dispersion relation $\omega^2=gk\tanh(kh_0)$. The colorbar is $\log_{10}(E)$, $E$ in m$^4$s.}
\end{figure*}

To get more insight into the changes of spectra in terms of $T_p$, it is worth analysing the frequency-wave vector spectra $E(k_x,k_y,\omega)$. Cuts of this spectrum for $k_y=0$, for $k_x=0$ and for $\omega/2\pi=1.5$~Hz are shown in figs.~\ref{spkxf35}, \ref{spkyf35} and \ref{spkxky35} respectively. Let us focus first on the cut at $k_y=0$ (Fig.~\ref{spkxf35}). At $T_p=1$~s, most of the energy of the wave field lies on lines that correspond to free dispersive waves along the linear dispersion relation of the dispersive waves at finite depth (black solid line, eq. (\ref{eq-rd})). The energy is maximum at the forcing frequency $(\pm k_p,\omega_p)$ and decays rapidly when moving away from the peak. The increase of $T_p$ to $1.5$~s and $2$~s produces energy on extra ridges in addition to the free waves; these ridges are the signature of the so-called ``bound waves'' (see discussion on bound waves in appendix \ref{sec.bound}). Bound waves result from second order interactions between free waves. Their increased presence shows that quadratic nonlinear interactions strengthen as $T_p$ increases. In a deep water interpretation, the strength of the nonlinearity can be quantified by $k_p\eta_{rms}$ which is actually decaying when $T_p$ increases. Thus the reinforcement of the bound waves is consistent with increasing nonlinearity associated with finite depth effects. Note also that more and more energy appears at low frequencies below $f_p$ as observed in the frequency spectra, which is also consistent with the reinforcement of second order interactions. 

At even higher $T_p\geq 3$~s, the ridges merge into a sort of continuum and some energy is also concentrated on the shallow water dispersion relation (black dashed line) and even possibly along a straight line with a velocity which is slightly larger than $c_0$. In Fig.~\ref{spkyf35} (cut of the spectrum for $k_x=0$), the energy is confined on the free wave dispersion relation in all cases. This suggest a strong anisotropy, particularly at large $T_p$. 

The increase in anisotropy is confirmed in Fig.~\ref{spkxky35} that shows a cut of the spectrum at the frequency $\omega/2\pi=1.5$~Hz. At $T_p=1$~s, the energy distribution is rather isotropic although the forcing with $s_{max}=30$ has a rather strong directivity along the $x$ axis (angular width at $f_p$ equal to $\pm 17^\circ$). It means that the non linear dynamics operates an angular redistribution of the energy while the waves bounce back and forth in the wave tank. When increasing the value of $T_p$, the energy distribution along the free wave dispersion relation (outer dashed circle) becomes more and more anisotropic. Bound waves appear mostly in the vicinity of the $k_x$ axis as the most energetic free waves are also propagating in these directions. At $T_p=5$~s and $T_p=10$~s the distribution of the energy changes drastically. The energy of the free waves decrease very significantly. Most of the energy is focused on two red dots on the $k_x$ axis and on the shallow water dispersion relation (inner dashed circle). At $T_p=10$~s, the red dots actually lie slightly inside the inner dashed circle, meaning that these energetic waves propagate at a velocity slightly faster than $c_0$. The observation of energy localized mostly on a line $\omega=c k$ with $c$ a little larger than $c_0$ strongly suggests the presence of solitons. Indeed, as a soliton propagates without changing shape, all its Fourier components propagate at the same velocity $c$ so its energy spreads on one line in a $(k,\omega)$ spectrum. Furthermore, solitons propagate at a velocity slightly larger than $c_0$ (as discussed in the introduction) which is consistent with our observations. The important conclusion from these observations is that for $T_p\geq 3$~s, solitons are present while they are more likely not to be present for lower values of $T_p$. From table~\ref{table_Tp}, $T_p\geq 3$~s corresponds to $Ur$ larger than 0.25 which is consistent with the regimes identified in \cite{zhao2024}  suggesting that regimes with $Ur>0.25$ correspond to cnoidal waves (which includes solitons).

\begin{figure}[!htb]
\includegraphics[width=16cm]{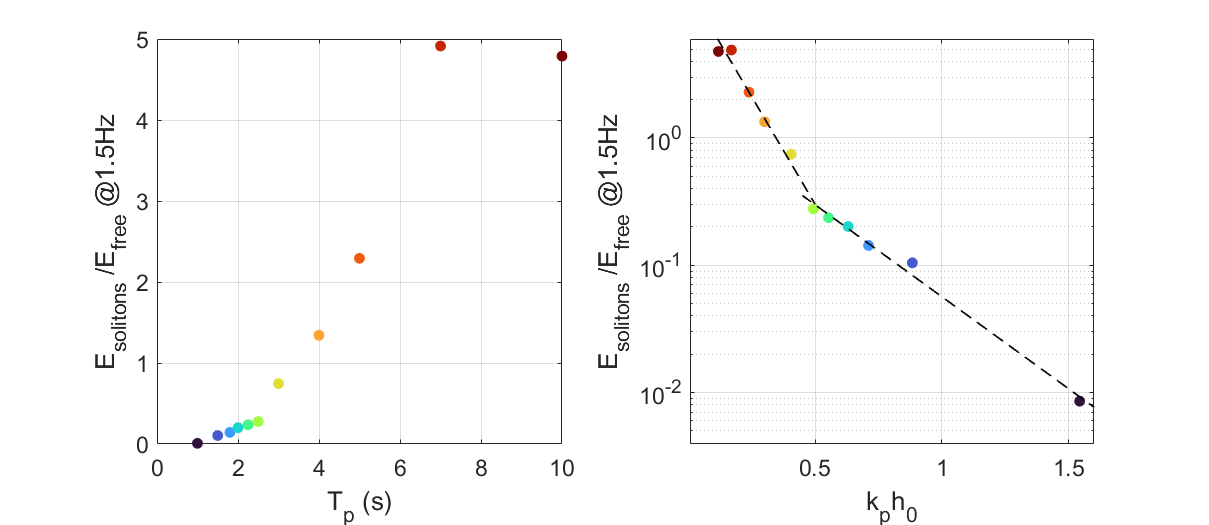}
\caption{\label{fig:ratio} Evolution of the ratio $E_{solitons}/E_{free}$ at $1.5$~Hz for $H_{m0}=3.5$~cm (see text for definition). Color symbols are the same as the colors of the curves in Fig.~\ref{spf35}. Left: ratio as a function of $T_p$. (b) ratio as a function of the dispersion parameter $k_ph_0$. Dashed lines are linear fits of the log of the ratio as guides for the eye.}
\end{figure}

In summary, we observe a change in wave regime between weak turbulence at low $T_p$ and solitons at large $T_p$. Next, we would like to quantitatively investigate the transition between these regimes. Due to the finite resolution of our measurements in the spectral space, disentangling energy associated to waves following the finite depth dispersion relation from the waves associated to shallow water dispersion relation is difficult. The various branches are clearly separated only at frequencies higher than $1$~Hz. In order to make a quantitative distinction between experiments with solitons and experiments without solitons, we chose to compare the energy associated to each dispersion relation at the frequency $1.5$~Hz.
Moreover, since the solitons in our context primarily travel along directions close to the $x$-axis, we limit the comparison to directions forming angles with the $x$-axis of less than $\pi/4$ in absolute value, whether positive or negative.
Within these angular ranges, we integrate the energy contained along the white circles shown in Fig.~\ref{spkxky35}, considering a thickness of $\delta k = \frac{2\pi}{L_{meas}}$ on either side of each circle (with $L_{meas}$ the size of the measurement region of the stereoscopic system, $\delta k$ is the resolution of the Fourier transform).

For the soliton branch we allow for an additional $\delta k$ inside the circle as the solitons propagate slightly faster than $c_0$. 
We define the ratio of the two energies as $E_{solitons}/E_{free}$ and present the result in Fig.~\ref{fig:ratio}.
The ratio is observed to increase significantly with increasing $T_p$ or decreasing dispersion factor $k_ph_0$. A clear change in slope appears, with a transition occurring when the ratio approaches 0.3 (corresponding to $T_p = 2.5\,$s  for the selected dataset). This transition is interpreted as a shift between two distinct regimes: one dominated by solitons when $E_{solitons}/E_{free} > 0.3$, and another dominated by dispersive waves when $E_{solitons}/E_{free} < 0.3$. This quantitative estimate is fully consistent with the visual analysis of the spectra shown in Fig.~\ref{spkxf35}: experiments with $E_{solitons}/E_{free} > 0.3$ exhibit a well-defined energy concentration near the shallow water dispersion relation. Accordingly, we adopt $E_{solitons}/E_{free} = 0.3$ as a threshold to uniformly classify all experiments in the subsequent analysis (see below).

\subsection{Distribution of water surface displacements}

\begin{figure}[!htb]
(a)\includegraphics[width=12cm]{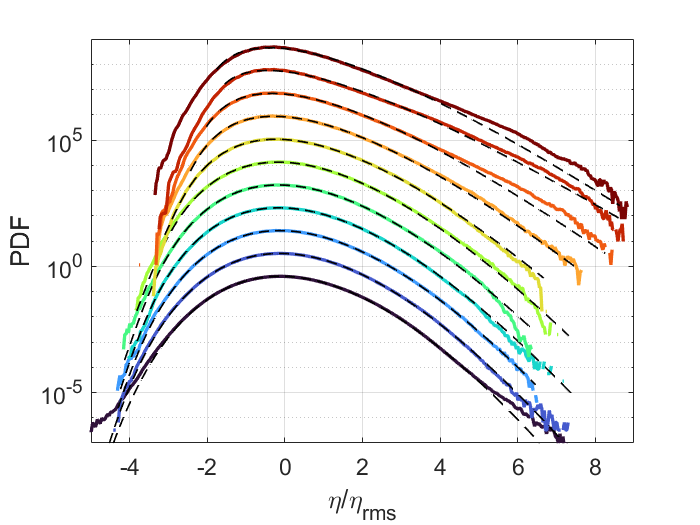}

(b)\includegraphics[width=12cm]{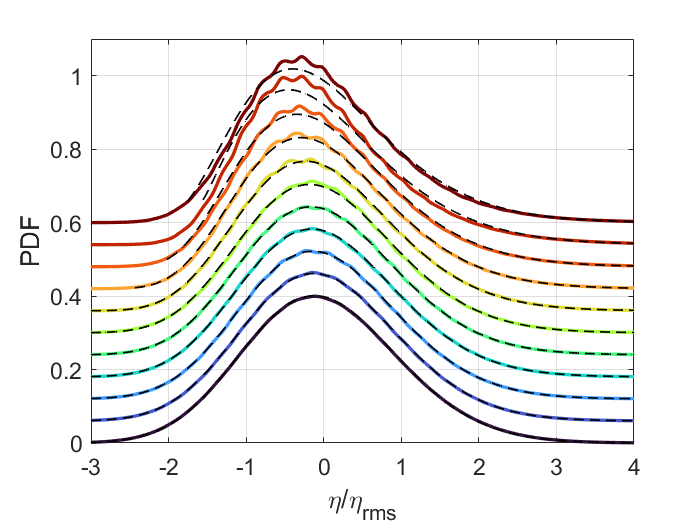}
\caption{\label{fig:PDF}PDF of the water surface displacement for the experiments of Table~\ref{table_Tp}. Curves have been shifted vertically for clarity. $T_p$ is increasing from bottom to top (from 1~s to 10~s, same color code as in Fig.~\ref{spf35}). (a) semilog scale and (b) linear scale. Black dashed lines are Tayfun type distributions (\ref{eq:Tayfun}) computed for each experiment based on the value of the measured skewness $S$ (see text for details).}
\end{figure}

Distributions of the water surface displacement are shown in Fig.~\ref{fig:PDF}. The small oscillations of the PDF near its peak are due to a small ``peak locking'' effect which is well known for image correlation techniques such as Particle Image Velocimetry due to pixel discretization. These oscillations are non-physical and are solely artifacts from image processing.  
As $T_p$ increases, the distribution becomes increasingly asymmetric, with a pronounced tail emerging on the side of positive elevations. Such asymmetry is characteristic of quadratic nonlinearities, and it is well established that nonlinear waves tend to exhibit sharp crests and flat troughs. In deep water, Tayfun proposed a model for such distributions that incorporates quadratic nonlinearities~\cite{Tayfun}. The model is based on the assumption of a narrow-banded wave spectrum and relies on a Stokes-type expansion. Socquet-Juglard {et al.} proposed a simplified expression~\cite{Socquet}:
\begin{equation}
P(\eta)=\dfrac{1}{\pi\sigma}\int_0^\infty\exp\left[-\dfrac{x^2+(1-C(x))^2}{2\sigma^2}\right]\dfrac{dx}{C(x)}
\label{eq:Tayfun}
\end{equation}  
with $C(x)=\sqrt{1+2\sigma \eta+x^2}$ (note that $P(\eta)$ is defined only for $\eta>-\dfrac{1}{2\sigma}$) and $\sigma/\sqrt{2}$ is the deep water steepness of the wave. 
In finite depth, the interpretation of $\sigma$ as a nonlinearity parameter is no longer valid. However, one can exploit the known relationship in deep water between $\sigma$ and the skewness factor, $S = \langle \eta^3 \rangle / \langle \eta^2 \rangle^{3/2} = 3\sigma$, to estimate an effective value of $\sigma$ based on the measured skewness. The distribution (\ref{eq:Tayfun}) is shown as black dashed lines in Fig.~\ref{fig:PDF} using the experimental value of the skewness. 
For the lowest values of $T_p$, the observed PDFs show excellent agreement with the Tayfun distribution. In the case of $T_p = 1$~s, the agreement is not as good for the most extreme values, which can be attributed to limitations in the stereoscopic reconstruction algorithm. This reduced performance is likely due to the relatively steep wave profiles present in this dataset.
In deep water, Onorato et al.~\cite{onorato09} also observed that, for wave fields with similar spectral directionality, the distribution of surface elevation closely followed the Tayfun distribution. Deviations from this model were reported only under highly directional forcing, where broader tails emerged due to modulational instability. This mechanism is not relevant in our case, as the wave directionality is moderate and the spectra are not narrow-banded.
When $T_p\geq 5$~s (three uppermost curves), the Tayfun distribution underestimates the probability of occurrence of the highest water surface displacements.
This is consistent with previous observations indicating a change in the nature of the wave field. It should also be noted that the narrow-band spectrum assumption underlying the Tayfun distribution is no longer valid for these experiments, as discussed earlier. Moreover, the Tayfun distribution fails to accurately capture both the peak of the PDF and the decay on the side of negative elevations.

\begin{figure}[!htb]
\includegraphics[width=10cm]{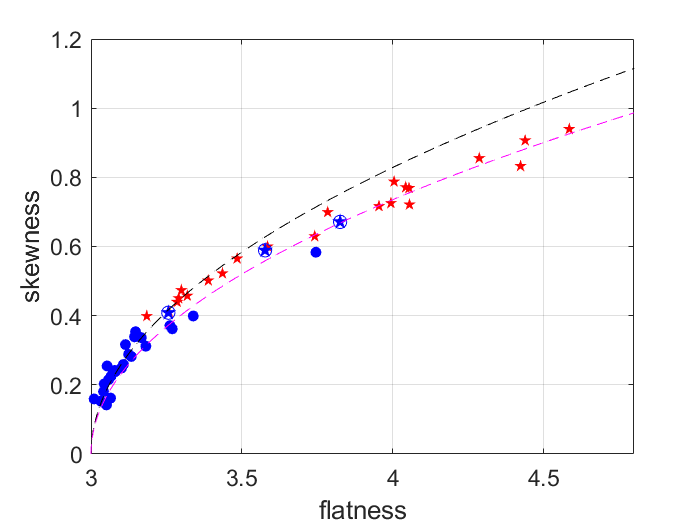}
\caption{\label{fig:skewflat}Skewness vs flatness for all experiments with $s_{max}=30$. Blue dots are experiments with $E_{solitons}/E_{free}<0.3$ (no solitons) and red stars are experiments with $E_{solitons}/E_{free}>0.3$ (with solitons). Blue circles with stars are non solitonic states for which a visual inspection of the $(\omega,k_x)$ spectra suggests that a few solitons may have existed but very scarce although the energy criterion classified these experiments as without solitons. Upper black dashed line is the relation for the Tayfun distribution, the lower magenta dashed line is the relation for a lognormal distribution ($F=3+1.85 S^2$) (both relations from \cite{zhang2024}).}
\end{figure}

\begin{figure}[!htb]
\includegraphics[width=16cm]{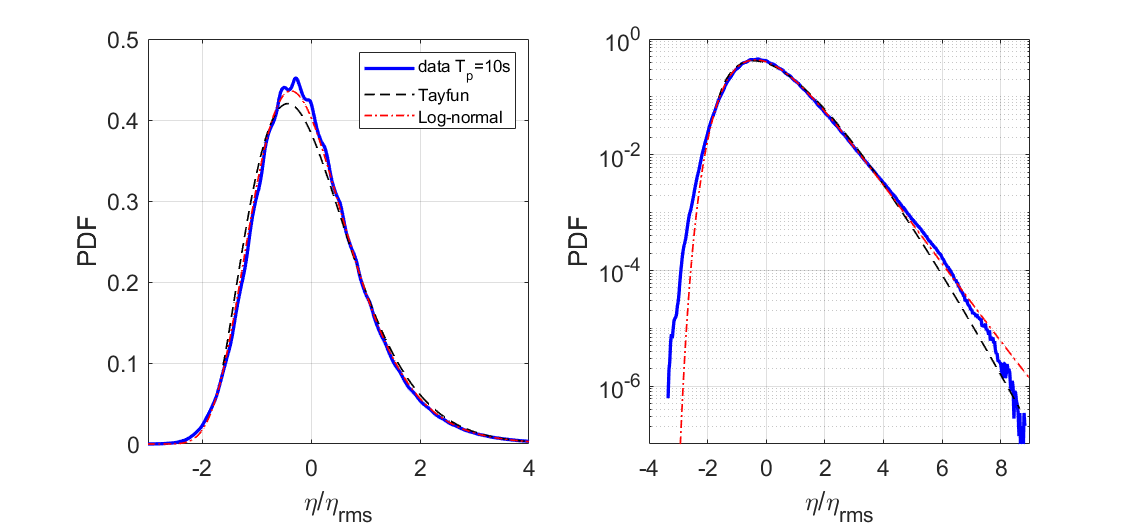}
\caption{\label{fig:lognPDF} water surface displacement PDF for the case $T_p=10$~s of Table~\ref{table_Tp}. The black dashed line is the Tayfun distribution using the experimental value of the skewness $S$ 
(\ref{eq:Tayfun}). 
The red dashed curve is the lognormal distribution using the experimental value of the skewness $S$.}
\end{figure}

A useful approach to quantify the evolution of the distribution’s shape is to compute the skewness $S$ and the flatness $F = \langle \eta^4 \rangle / \langle \eta^2 \rangle^2$, and to plot one as a function of the other, as shown in Fig.~\ref{fig:skewflat}. Each dot correspond to a experiment similar to those in table~\ref{table_Tp} but varying the amplitude $H_{m0}$ of the forcing as well (blue dots in fig~\ref{fig:params}). The dots spread along a line in this $(F,S)$ representation. Some scatter is observed that provides an estimate of the uncertainty of the estimation. The $(F,S)$ curve for the Tayfun distribution (taken from \cite{zhang2024,Tayfun2020}) is shown as the black dashed line. It remains close to observations for $F<3.5$ and $S<0.6$ but deviates beyond this limit. Zhang {\it et al.} \cite{zhang2024} suggested that the distribution of shallow water waves may be better described by a lognormal distribution (shifted to ensure that $\langle \eta \rangle=0$). The lognormal distribution gives $F =3+1.85 \, S^2$ \cite{zhang2024} and this relation is shown as the magenta dashed line in Fig.~\ref{fig:skewflat}. The experimental points seem to follow this law for $F>3.7$ that correspond to solitonic regimes. 
At lower values, the scatter in the data points does not allow a clear determination of which model provides a better fit to the data.
Fig.~\ref{fig:lognPDF} shows a comparison of the distribution for the experiment with $T_p=10$~s in table~\ref{table_Tp} compared with the Tayfun and lognormal distribution. 
The parameters of both distributions are computed from the estimated skewness. Although the differences are small, they are noticeable and indicate that the lognormal distribution provides a better fit to the experimental data, both near the peak and for the tail of positive $\eta$. The agreement is less satisfactory for extreme troughs, and a comparison with the Tayfun distribution is not feasible for large negative values.

\subsection{Phase diagram}

\begin{figure}[!htb]
(a)\includegraphics[width=8cm]{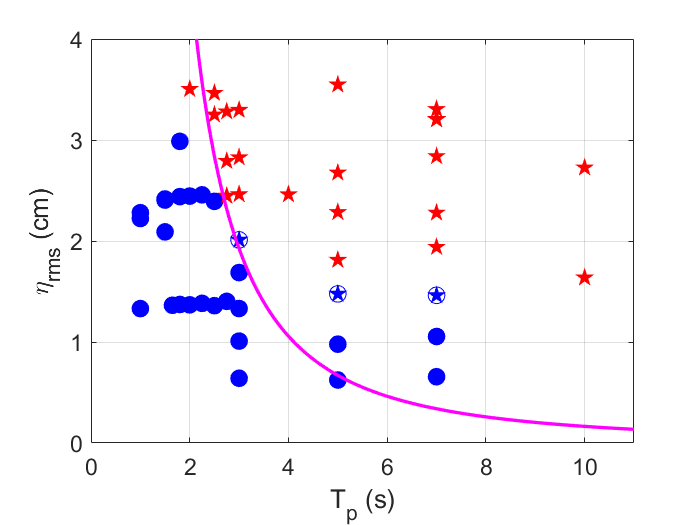}
(b)\includegraphics[width=8cm]{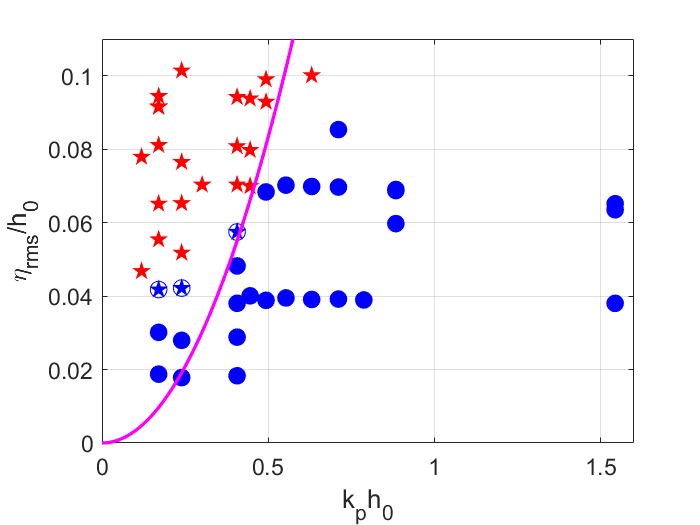}

(c)\includegraphics[width=8cm]{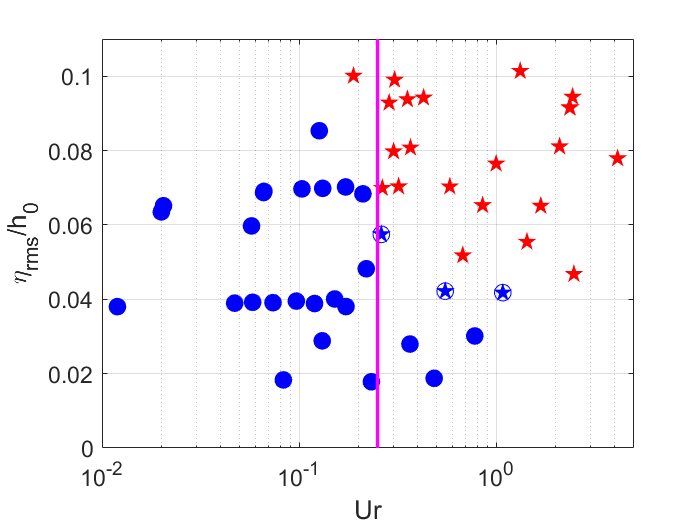}
(d)\includegraphics[width=8cm]{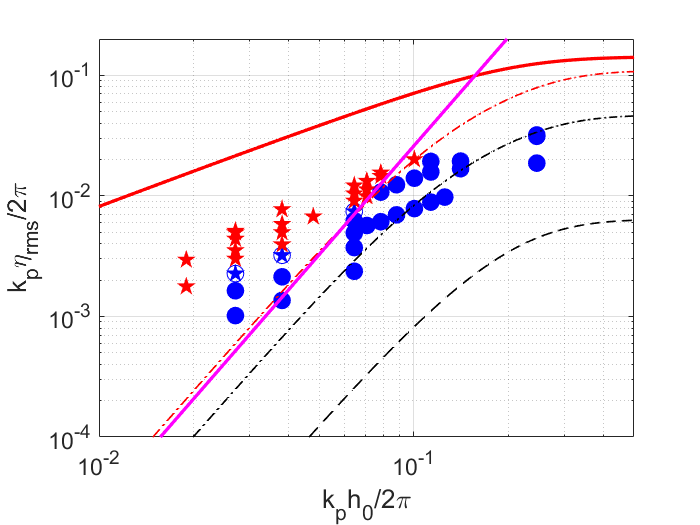}
\caption{\label{fig:phasediag}Phase diagram of JONSWAP experiments with $s_{max}=30$ using several sets of parameters. (a) as a function of $\eta_{rms}$ vs $T_p$, (b) $\eta_{rms}/h_0$ vs $k_ph_0$, (c) $\eta_{rms}/h_0$ vs $Ur$, (d) representation {\it à la Le Méhauté} from \cite{lemehaute,zhao2024}: $k_p\eta_{rms}$ vs $k_ph_0$. Blue dots are non solitonic state and red stars are solitonic states. Blue circles with stars are non solitonic states for which a visual inspection of the $(\omega,k_x)$ spectra suggests that a few solitons may have existed but were very scarce although the energy criterion classified these experiments as without solitons. These states are transitional. Magenta line: $Ur=0.25$, it corresponds to the limit where the development in Stokes waves is not relevant and cnoidal wave theory must be considered \cite{zhao2024}. Black dashed line: limit at which the second order term in the Stokes development is 1\% of the linear term (limit of validity of the linear theory). Black mixed dashed line: limit at which the third order term is 1\% of the first and second order terms (limit of second order theory). Red mixed dashed line, same but with 5\%. Red solid line: limit of wave breaking in free water (far from walls where amplification of the wave amplitude can happen).}
\end{figure} 

We use the criterion $E_{solitons}/E_{free}>0.3$ for all experiments with $s_{max}=30$ as above to define a phase diagram of our experiments that is shown in Fig.~\ref{fig:phasediag}. By ``phase diagram'' we mean a classification of the solitonic vs non solitonic states as a function of the physical parameters (several choices are possible for the parameters and some of them are shown in the figure). We also did a systematic visual inspection of the $(k_x,\omega)$ spectra to ensure that energy along the line or near the shallow water dispersion relation $\omega=c_0k$ was present. Although this visual detection is somewhat subjective, the two methods were in agreement except for 3 experiments shown as blue stars with circles that lie near the border between the two regimes. One can define an Ursell number as, 
\begin{equation}
Ur=\frac{3k_p\eta_{rms}}{4(kh_0)^3} \, .
\end{equation}
As discussed in \cite{zhao2024}, the nonlinear regime is expected to change for $Ur\approx 0.25$. The magenta line in Fig.~\ref{fig:phasediag} corresponds to $Ur=0.25$. We see that for large values of $\eta_{rms}$, this line is rather consistent with our observations but not for smaller values ($\eta_{rms}\lesssim 1$~cm). Several reasons can be invoked to understand this discrepancy that concern either finite size effects or detection issues:
\begin{itemize}
\item For $T_p=5$ and $7$~s, the wavelength of the waves is $9.2$ and $12.9$~m (resp.) which are not small compared to the size of the wave tank that is $27$~m in the $x$ direction. Thus finite effects may impact the propagation of the dispersive waves and the solitons by resonances in the tank. 
\item The generation of solitons occurs through a fission mechanism as described by Zabuski \& Kruskal~\cite{ZK}: a sine steepens by nonlinearity to form a shock that is ultimately regularized by dispersion. For an initial sine wave of frequency $\omega$ and amplitude $\eta_0$, the distance of the formation of the shock is $L_b=\dfrac{2c_0h_0}{3\omega\eta_0}$ \cite{Trillo}. For $\eta_0\approx \eta_{rms}\approx 1$~cm and $T_p=5$ or $7$~s, the breaking distance is 34 and 48~m respectively. This distance is larger than the size of the tank thus one expects finite size effects by nonlinear interaction with the walls, with the wavemakers and between reflected waves before the shock formation occurs. 
\item At such small amplitudes, the number of solitons generated per period of the sine oscillation is typically one (following \cite{Trillo}) which means that few solitons are generated during the 20 minutes of recording. Thus, there may also be  a lack of statistics that make the detection of solitons more difficult because of their scarcity. 
\item The small amplitude of the solitons also makes them hard to differentiate from dispersive waves using our criteria. The adequate tool for the detection of soliton is of course the Inverse Scattering Transform but we are not aware of any implementation of a practical algorithm for periodic boundary conditions (or even in the infinite case actually) that would allow us to implement such a detection in practice (as is the case in 1D for instance, see \cite{Trillo,RedorPRF} for examples).
\end{itemize}

In Fig.~\ref{fig:phasediag}(d), one can see that most non solitonic states are close to the limit of validity of the second order theory (but below the limit) which explains that the wave surface displacement PDF is well described by the Tayfun distribution. Conversely, the solitonic states are in the domain of validity of cnoidal waves, which is again consistent with the presence of solitons. Again, the experiments corresponding to $T_p=5$ and $7$~s with low amplitude (four blue dots at bottom left part of Fig.~\ref{fig:phasediag}(d)) appear inconsistent with the global picture.

\subsection{Effect of the finite size of the tank on the directivity of the waves}
\begin{figure}[!htb]
\includegraphics[width=14cm]{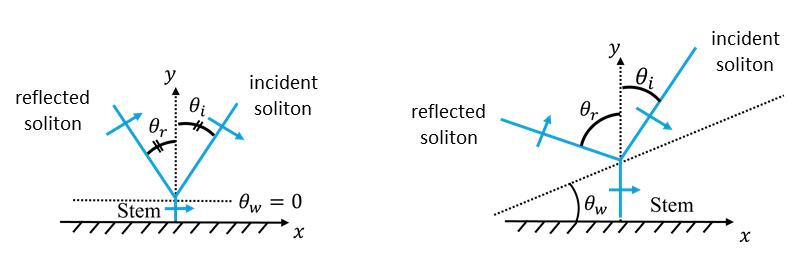}
\caption{\label{fig:stem}Schematics of the reflection laws of a line soliton on a wall. The left one represents a regular reflection ($\kappa>1$) and the right one shows the case of Mach reflection ($\kappa<1$). Arrows indicate the propagation direction of each part of the wave. See text for details.}
\end{figure}

One open question is the reason of the extreme directivity of the solitons that propagate along the $x$ direction while the forcing is not as focused in directivity. One reason is related to the fact that reflection laws of solitons on the lateral wall are much more complex than for linear waves. In particular they do not follow the usual Snell-Descartes laws. Miles did some early work on the classification of soliton interactions for the KdV solitons \cite{Miles77a,Miles77b}. The work of Miles has been extended to the KP solitons \cite{Jia} for small angles and by Kodama \& Yeh \cite{KodamaYeh} for an extended KP case. They showed that the properties of the reflection depend on the parameter $\kappa$ defined as
\begin{equation}
\label{eq:kappa}
\kappa=\dfrac{\sqrt{1+\sqrt{1+5\epsilon_i}}\tan \theta_i}{\sqrt{6\epsilon_i}\cos \theta_i}
\end{equation} 
where $\theta_i$ is the angle of the crest of the incident soliton with the normal to the wall and $\epsilon_i$ is the amplitude of the incident soliton normalized by $h_0$. The reflection is characterized by the presence of a stem, i.e. a wave that is normal to the wall over some distance and that connects to the incident and reflected solitons (see Fig.~\ref{fig:stem}). Depending on whether $\kappa$ is larger or smaller than 1, the reflections, and in particular the dynamics of the stem, are different. {For $\kappa>1$, the reflection is regular with equal angles and amplitudes of the incident and reflected waves. Moreover the stem has a short crest length (comparable to the width of the soliton) and constant in time as the soliton propagates along the wall.} In the case $\kappa<1$ which occurs at small incidence and finite amplitude of the soliton, the situation is very different (Mach reflection). The angle $\theta_r$ of the reflected soliton is significantly larger than that of the incident soliton. Furthermore, the amplitude of the reflected soliton is significantly smaller than that of the incident soliton. Finally the length of the stem increases linearly with time as it propagates along the wall. It means that after a long time the stem will have a length comparable to that of the tank (see Appendix \ref{app:Mach} for an example). When the directivity of the forcing is relatively narrow, then $\theta_i$ will be small for the reflections taking place on the walls at $y=0$ and $y=30$~m and most likely $\kappa$ will be smaller than one for most detectable solitons for which $\epsilon_i$ is not small. It means that most of these solitons will develop a long stem and a small reflected soliton so that most of the energy of the solitons {leak into} the stems that are perpendicular to the wall. The presence of the walls make the propagation of the solitons almost solely along the $x$ direction. For $\kappa<1$ the stem is a true soliton and its amplitude is larger than that of the incident soliton so that a stem has a strong signature in the spectra in the soliton branch. 
Note that for these solitons the reflection on the walls at $x=0$ (wavemakers) and $x=27$~m is regular as the incident angle is close to normal.

\begin{table}[!htb]
\begin{ruledtabular}
{\begin{tabular}{lcc}
$s_{max}$	& 30 & 1\\
$\eta_{rms}$ & 3.2 & 3.3\\
$Ur$ & 2.36 & 2.45\\
$S$ & 0.88 & 0.69 \\
$F$ & 4.4 & 3.9 \\
$E_{solitons}/E_{free}$ & 2.0 & 1.4\\
\end{tabular}}
\end{ruledtabular}
\caption{\label{table:sp1}Parameters for the experimental runs with $T_p=7$~s and $H_{m0}=4$~cm and two different directivities with $s_{max}=1$ and $s_{max}=30$.}
\end{table}

\begin{figure}[!htb]
(a)\includegraphics[width=8cm]{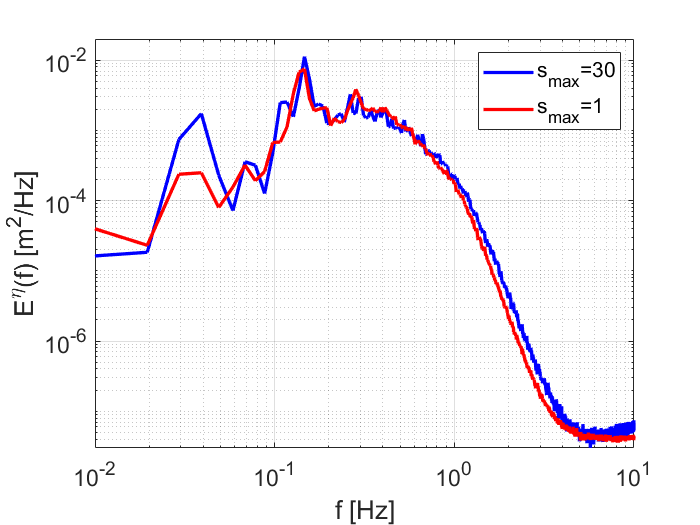}
(b)\includegraphics[width=8cm]{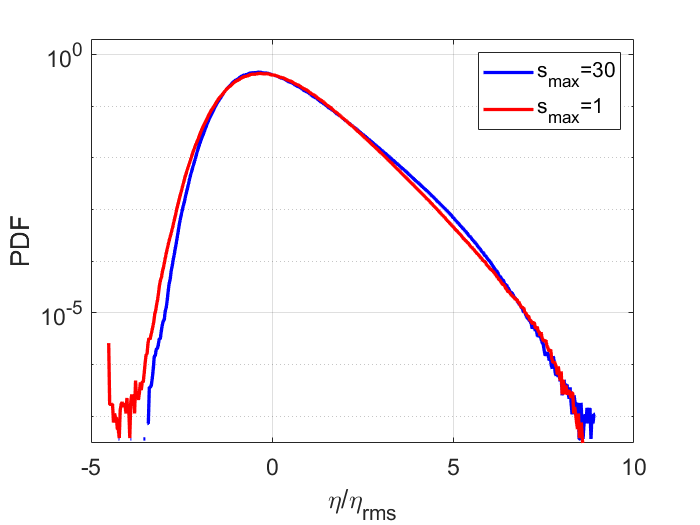}
\caption{\label{fig:sp1spectref}
(a) Comparison of water surface displacement power spectra of experiments for $T_p=7$~s and $H_{m0}=4$~cm for two directivities $s_{max}=30$ (blue, directional) and $s_{max}=1$ (red, almost omnidirectional). (b) Comparison of PDFs (same colors).}
\end{figure} 

\begin{figure}[!htb]
\includegraphics[width=14cm]{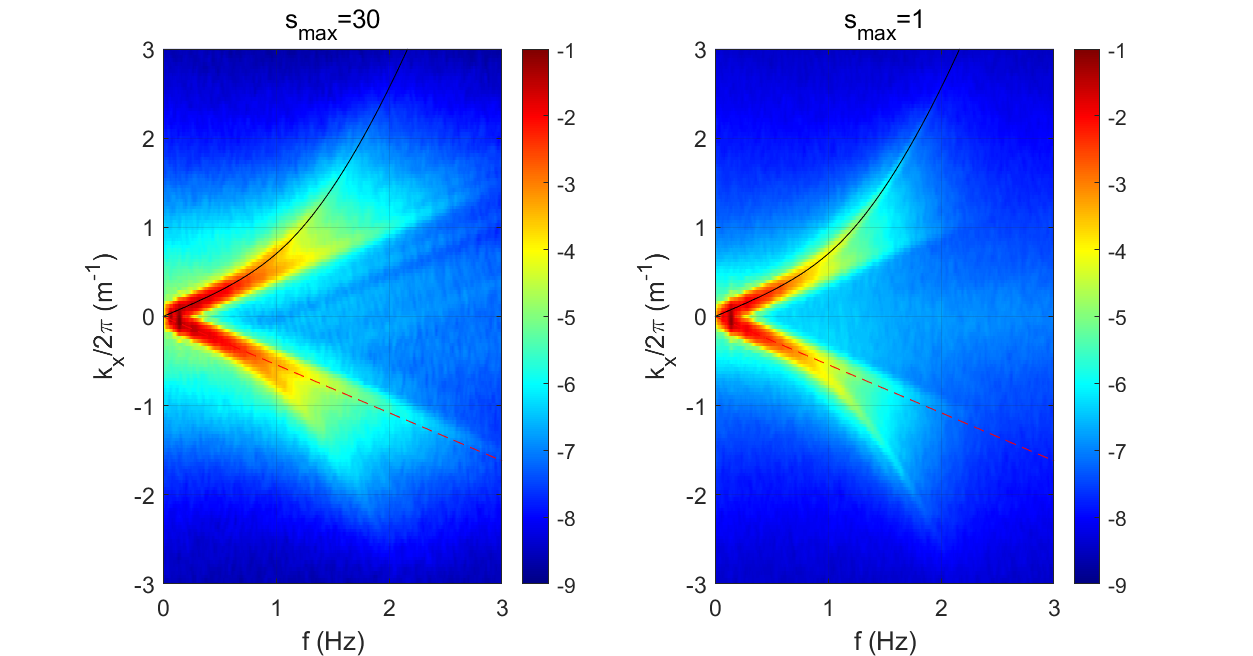}
\includegraphics[width=14cm]{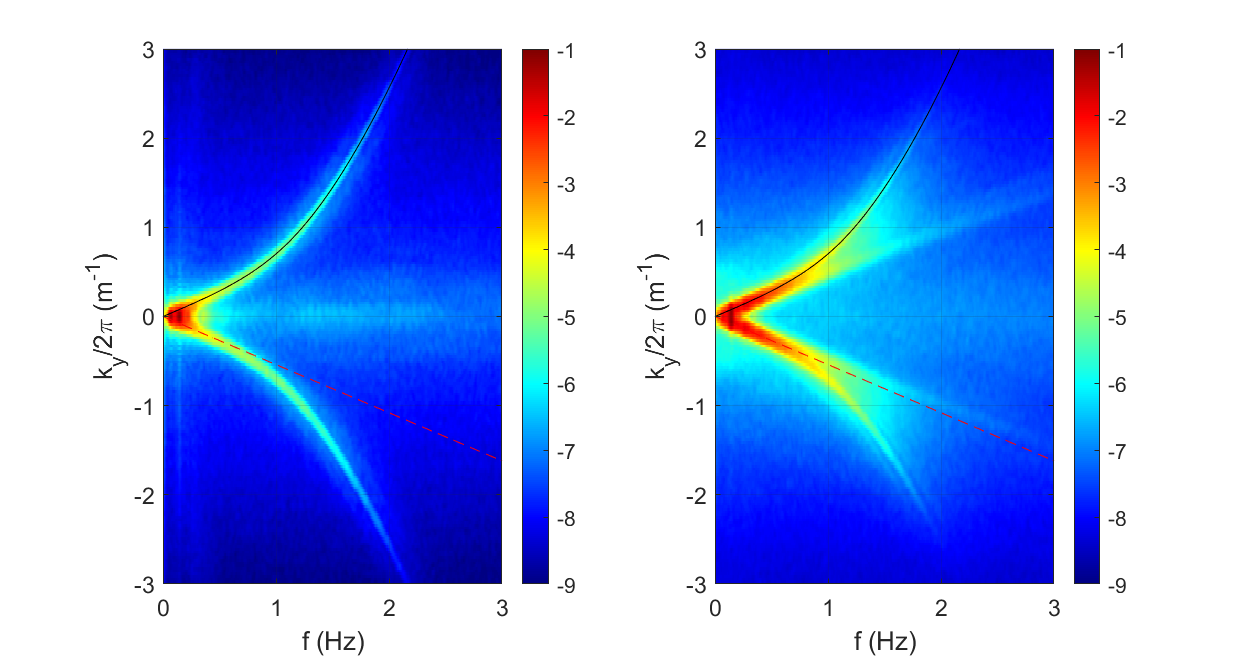}
\includegraphics[width=14cm]{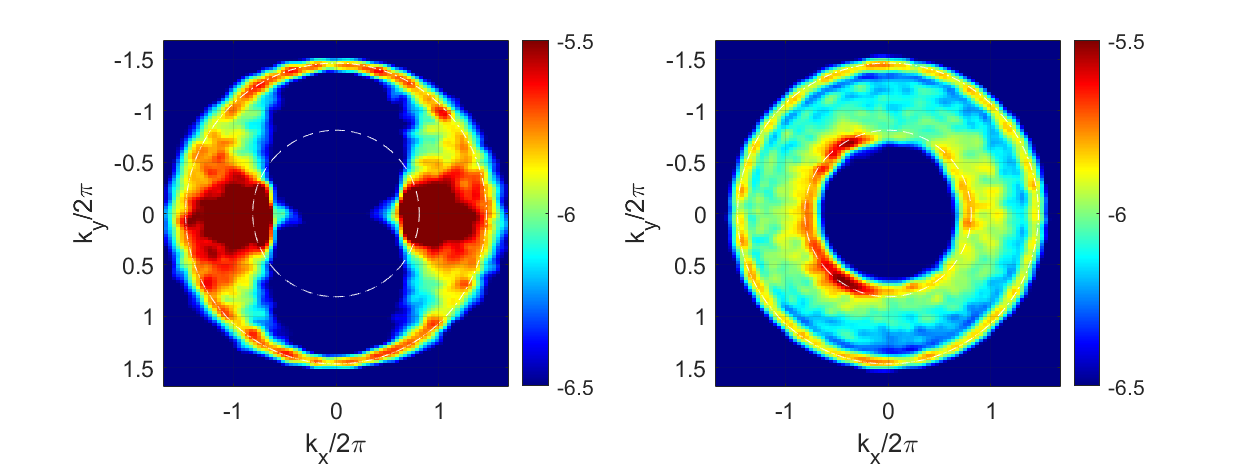}
\caption{\label{fig:sp1spectrekf}
Comparison of $(k,\omega)$ spectra of experiments for $T_p=7$s and $H_{m0}=4$cm when changing the directivity through $s_{max}$ from 30 (left) to 1 (right). Top line: $E(k_x,k_y=0,\omega)$. Middle line: $E(k_x=0,k_y,\omega)$. Bottom line: $E(k_x,k_y,\omega=3\pi)$, cut at $1.5$~Hz.}
\end{figure}

With these reflection properties in mind, additional experiments were carried out using a broad directional spreading of the forcing ($s_{max}=1$), a peak period of $T_p = 7$~s, and large wave amplitudes, in order to operate within the solitonic regime(see table~\ref{table:sp1} for the parameters). Averaging over four experiments was required to ensure a sufficient number of solitons propagating in all directions. As the incident angle can be much larger in that case, many solitons can have a value of $\kappa$ that is larger than 1. The statistics are compared with a similar experiment with $s_{max}=30$ in figs.~\ref{fig:sp1spectref} and \ref{fig:sp1spectrekf}. The frequency spectra and distribution of water surface displacement are quite similar, with small differences even though the {\it rms} of the wave surface displacements are almost the same, and thus the values of the Ursell number are close. However, the $(\mathbf k,\omega)$ spectra show a very different behavior in terms of directivity of the solitons. At $s_{max}=1$ solitons' propagation directions scatter on a wide range of angles up to $\pm \pi/2$ (the solitons propagating along the wavemakers are impossible to generate directly by the wavemakers and are induced by the nonlinear interactions inside the wave tank). The energy of the dispersive waves is comparable in both cases. The energy of the solitons propagating away from the wavemakers (with negative $k_x$) is larger than the solitons in the opposite directions but solitons are visible in all directions. Note also that a continuum of weak bound waves is also visible in all directions while bound waves are mostly concentrated along the $x$ direction for the $s_{max}=30$ cases. At $s_{max}=1$, the soliton gas is quite isotropic but the amplitudes of the solitons are relatively weaker. This is most likely due to the fact that for $s_{max}=30$, solitons whose crest is almost parallel to the wall of wavemakers are stems that have a larger amplitude than the incident solitons. Furthermore, such solitons can be reinforced when reflecting on the wavemakers if the phase of the pistons is adequate (during a forward push at the same time as the soliton hits the wavemakers). Many such reinforcements were observed during the experiments. This process cannot really occur when the soliton crest has a large angle compared to the wavemakers. It seems also that solitons propagating with $\theta_i$ close to $\pi/4$ for $s_{max}=1$ are more energetic.
This may be due to the fact that solitons propagating along diagonals of the tank are quite stable under regular reflections and they somehow resonate geometrically in the tank.

\section{Concluding remarks}

In summary, we performed a parametric experimental study of the dynamics of wave turbulence with 2D propagation in a large wave tank. By systematically varying the dispersion and the nonlinearity of the forcing, through adjustments to the peak frequency $f_p$ and the significant wave height $H_{m0}$, we observed a clear transition in the statistical properties of the wave field from fields with no solitons to ones with solitons in addition to dispersive waves.
In the dispersive wave regime without solitons, reducing dispersion is observed to have a strong effect on angular redistribution of the wave energy from the anisotropic forcing. While the spectrum is rather isotropic for deep water, it remains much more directional when the Ursell number is increased. Solitons seem to appear when $Ur$ becomes larger than $0.25$ (for $\eta_{rms}/h0 \geq 0.04$), that is a commonly accepted limit. 
The Tayfun distribution, relevant for dispersive waves, fails to predict the increase of the probability of extreme values induced by the emergence of solitons for $Ur>0.25$. The Lognormal law is better suited to describe the solitonic regime but it lacks physical and theoretical justification so far.

Finite size effects also independently influence the system through wave reflection at the boundaries. Vertical walls were used to reflect the waves, allowing for prolonged interactions. In contrast, replacing the vertical walls with absorbing beaches would lead to rapid wave dissipation, significantly limiting nonlinear interactions in both the weak turbulence and solitonic regimes.
The presence of walls leads to significantly more complex reflection behavior for solitons. Nonlinearity plays a crucial role in these reflections, especially at small angles of incidence. As a result, when the forcing is highly directional along the $x$ axis, the pronounced development of stems at the walls enhances the solitons' directivity, causing them to propagate primarily along the $x$ direction. Isotropic spreading of solitons can only be expected under nearly omnidirectional forcing. It is worth noting that this finite size effect would not occur under periodic boundary conditions, as commonly employed in numerical simulations.

\begin{acknowledgments}
This study has received funding from Agence Nationale de la Recherche project SOGOOD (ANR-21-CE30-0061) and from the Simons Foundation through the Simons Collaboration on Wave Turbulence.
TL acknowledges the support of the French Ministry of Higher Education and Research through the PhD grant.
The authors are grateful to ARTELIA company for making their wave tank available and fully operational.
\end{acknowledgments}

\appendix

\section{JONSWAP spectrum \label{app:jonswap}}

        Hasselmann {\it et al.} \cite{jonswap}, through the analysis of data collected as part of the JOint North Sea WAve observation Project (JONSWAP), have parameterized a typical wave spectrum that can describe many sea states. In the shallow case the relevance of the spectrum is questionable, but it is a convenient tool for generating multidirectional random waves. 
        {The energy distribution 
        for the frequency part of the spectrum JONSWAP is:}

        \begin{equation}
            S(f)= \alpha H_{m0}^2 \frac{f_p^4}{f^5} e^{-\frac{5}{4}\left( \frac{f_p}{f} \right)^4} \gamma^\beta
            \label{eq:JONSWAP_S(f)}
        \end{equation}
        \begin{equation}
            \textrm{with } \beta=\exp{\left( -\frac{(f-f_p)^2}{2(\sigma f_p)^2}\right)}
            , 
            \qquad
            \alpha=\frac{0.064}{0.23+0.0336\gamma-\frac{0.185}{1.9+\gamma}}, \quad
              \sigma=\left\{ 
                \begin{matrix} 
                0.07, f\leq f_p \\ 0.09, f > f_p \end{matrix} 
                \right.
        \end{equation}
        
        where $H_{m0}$ is the significant wave height equal to four times the standard deviation of the surface elevation, $f_p$ is the peak frequency and $\gamma$ is a shape factor (the larger $\gamma$ is, the more pronounced the peak of the spectrum). 
        
        For angular distribution, we use $\cos ^{2s}$ type of distribution \cite{goda} as:

        \begin{equation}
            D(f,\theta)=\frac{\Gamma(s+1)}{2\sqrt{\pi}\Gamma(s+1/2)}\cos^{2s}\left(\frac{\theta-\theta_0}{2}\right)
            \label{eq:JONSWAP_D(f,theta)}
        \end{equation}
        with 
        \begin{equation}
            s=s_{max}(f/f_p)^5 \textrm{ for } f\leq f_p \quad , \quad s=s_{max}(f/f_p)^{-2.5} \textrm{ for } f > f_p \, ,
        \end{equation}
        %
{and  $\theta_0$ is the main direction of wave propagation   
(in the present experiments, $\theta_0=0$ is the direction perpendicular to the wavemakers).}
The value of $s_{max}$ gives the directivity of the distribution. 
        
To generate the signals to drive the 60 piston wavemakers, we first generate a Gaussian random noise $\eta_r(y_i,t)$ where $y_i$ is the center position of the $i$-th piston. A double Fourier transform provides $\eta(k_y,\omega)$ which is a complex random noise with random phases. Using the dispersion relation $\omega^2=gk\tanh(kh_0)$, we change variables of the spectrum $E(f,\theta)$ into $E(k_y,\omega)$. We multiply $\eta_r(k_y,\omega)$ by $\sqrt{E(k_y,\omega)}$ and do an inverse Fourier transform to get the target signals $\eta_t(y_i,t)$ for the water surface displacement on the wavemakers. These signals are then translated into displacements of the wavemakers.

\section{\label{sec.bound}Bound waves}

\begin{figure}[!htb]
\includegraphics[width=0.8\textwidth]{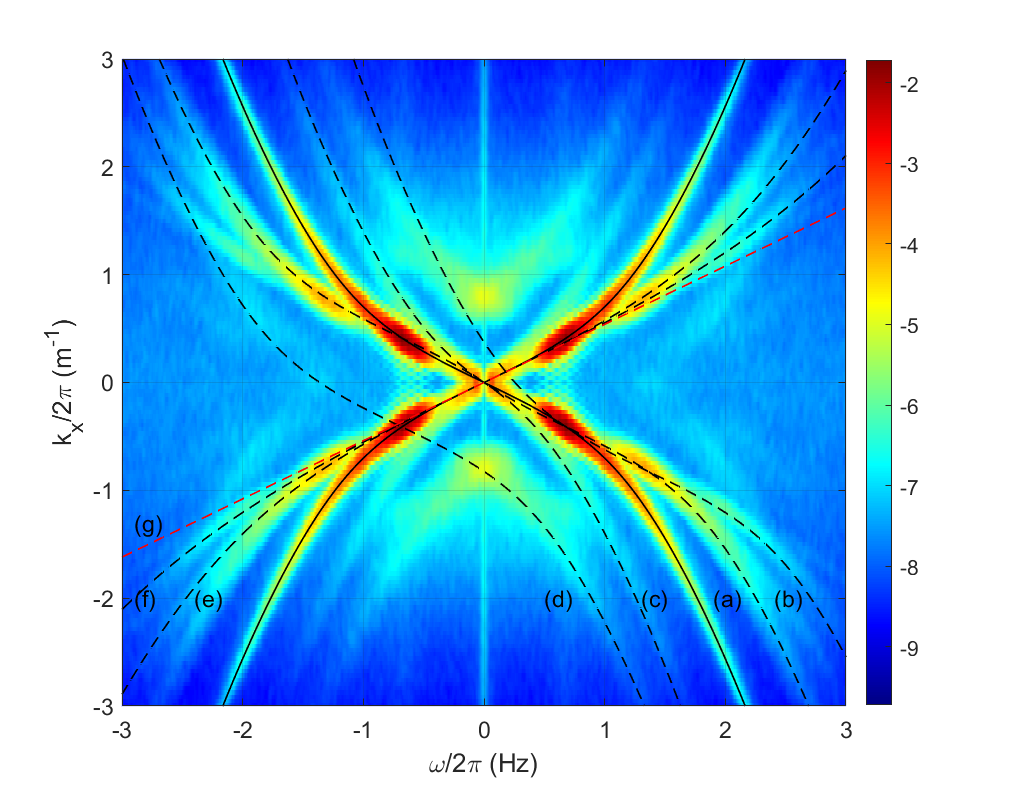}
\caption{\label{fig:bound}Bound waves in the spectrum $E(k_x,k_y=0,\omega)$ for the experiment with $H_{m0}=3.5$~cm and $T_p=1.5$~s (see table~\ref{table_Tp}). Solid black line: linear dispersion relation. Dashed line (a) is the linear dispersion relation $\omega^2=gk\tanh(kh_0)$ shifted by the vector $(\omega_p,-k_p)$
in the $(\omega,k_x)$ plane. Similarly (b) is shifted by $(2\omega_p,-2k_p)$,
(c) is shifted by $(-\omega_p,k_p)$
, (d) is shifted by $(-\omega_p,-k_p)$. (e) is the second harmonics and (f) is the third harmonics. Line (g), red dashed line, is the shallow water linear dispersion relation i.e. the soliton boundary $\omega=\sqrt{gh_0}k$. Only half the dashed lines have been shown to preserve some clarity in the figure: for all dashed lines, symmetric dispersion relations with respect to $\omega=0$ exist as well for waves propagating in the other direction.}.
\end{figure}

 \cite{toffoli2007} proposed an expression for the second order correction to the water surface displacement of linear waves. 
At linear level one can write the water surface displacement as:
\begin{equation}
\eta^{(1)}(\mathbf x,t)=\sum_{\mathbf k} a_{\mathbf k}\cos\left(\mathbf k\cdot \mathbf x - \omega_{\mathbf k}t+\phi_{\mathbf k}\right)
\end{equation}
where $\mathbf k$ are the wave vectors, $\omega_{\mathbf k}$ the associated frequencies following the linear dispersion relation, $\phi_{\mathbf k}$ are the phases and $a_{\mathbf k}$ are the amplitudes.

Then the second order correction reads:
\begin{equation}
\eta^{(2)}(\mathbf x,t)=\dfrac{1}{4}\sum_{\mathbf k_1\mathbf k_2}a_{\mathbf k_1}a_{\mathbf k_2}\left[K_{\mathbf k_1\mathbf k_2}^-\cos(\Phi_{\mathbf k_1}-\Phi_{\mathbf k_2})+K_{\mathbf k_1\mathbf k_2}^+\cos(\Phi_{\mathbf k_1}+\Phi_{\mathbf k_2})\right]
\end{equation}
where $\Phi_{\mathbf k_i}=\mathbf k_i\cdot \mathbf x - \omega_{\mathbf k_i}t+\phi_{\mathbf k_i}$. The expressions of $K_{\mathbf k_1\mathbf k_2}^-$ and $K_{\mathbf k_1\mathbf k_2}^+$ are given in \cite{toffoli2007}.
The term $\Phi_{\mathbf k_1}+\Phi_{\mathbf k_2}=(\mathbf k_1+\mathbf k_2)\cdot \mathbf x - (\omega_{\mathbf k_1}+\omega_{\mathbf k_2})t+\phi_{\mathbf k_1}+\phi_{\mathbf k_2}$ appears as a wave phase with $\mathbf k=\mathbf k_1+\mathbf k_2$ and $\omega=\omega_{\mathbf k_1}+\omega_{\mathbf k_2}$. Considering the curvature of the dispersion relation, $\mathbf k$ and $\omega$ built in this way never follow the linear dispersion relation. Thus from linear waves all following the linear dispersion relation, the second order effect sends energy in the Fourier space at positions $(\mathbf k,\omega)=(\mathbf k_1+\mathbf k_2,\omega_{\mathbf k_1}+\omega_{\mathbf k_2})$ out of the dispersion relation. The same occurs due to $K_{\mathbf k_1\mathbf k_2}^-$ with energy going to $(\mathbf k,\omega)=(\mathbf k_1-\mathbf k_2,\omega_{\mathbf k_1}-\omega_{\mathbf k_2})$. These Fourier modes are called ``bound waves'' by contrast to ``free waves'' that follow the dispersion relation. However these waves do not propagate independently, they exist because free waves are present. 

Figure~\ref{fig:bound} shows an example of an experiment for which many specific bound waves are visible. The continuous black line is the linear dispersion relation that is the most visible. But some energy can be seen also on other lines (dashed black lines). Line (a) corresponds to the bound wave if wave 1 is a free wave and wave 2 corresponds to the peak of the spectrum with $(\omega_p,\mathbf{k}_p)$. One gets:
\begin{equation}
\omega=\omega_1+\omega_p \quad \textrm{and} \quad \mathbf k=\mathbf k_1+\mathbf k_p
\end{equation}
Thus these bound waves lie on a line in the $(\omega,k_x)$ plane which is a translation of the dispersion relation by $(\omega_p,-k_p)$. Line (b) is the same process of coupling with the peak of the spectrum iterated once more, so that the dispersion relation is translated by $(2\omega_p,-2k_p)$. Line (c) is similar with a translation $(-\omega_p,k_p)$ which amounts to coupling with the peak $(-\omega_p,k_p)$. Line (d) is similar with a translation $(-\omega_p,-k_p)$ which amounts to coupling with the peak $(-\omega_p,-k_p)$. Line (e) is another sort of bound wave which is simply the second harmonics of the dispersion relation ($\mathbf k_1=\mathbf k_2$), which amounts to a dilatation of a factor 2 of the linear dispersion. Similarly line (f) is the third harmonics which is a dilatation of a factor 3. The red dashed line is the shallow water dispersion relation $\omega=\sqrt{gh_0}k$. At a given value of $|k|$, all bound waves (except line (d)) lie in a region where the frequency is lower than $\sqrt{gh_0}k$. By constrast, if a soliton is present in the tank, its energy will show up at frequencies slightly higher than $\sqrt{gh_0}k$ in a region where bound waves are almost not present (line (d) is very faint in this region). Due to the finite resolution of the evaluation of the spectrum, in practice, some energy may spread slightly on the ``wrong'' side of the shallow water dispersion relation. Note that these specific lines are very visible in this case as the spectrum of free waves is strongly peaked. However more generally all free waves (not only that of the peak of the spectrum) on the surface of the free dispersion relation will generate energy in the Fourier plane by second order coupling but here it is hidden by the experimental noise. For experiments at large $T_p$ the spectrum is not peaked and the bound waves appear as a continuum in between the two dispersion relations as seen in figs. \ref{spkxf35}, \ref{spkyf35}, \ref{spkxky35} and \ref{fig:sp1spectrekf}.  

\section{\label{sec.compsp}Comparison between capacitive gauges and stereoscopic reconstruction}

\begin{figure*}[!htb]
(a)\includegraphics[width=\textwidth]{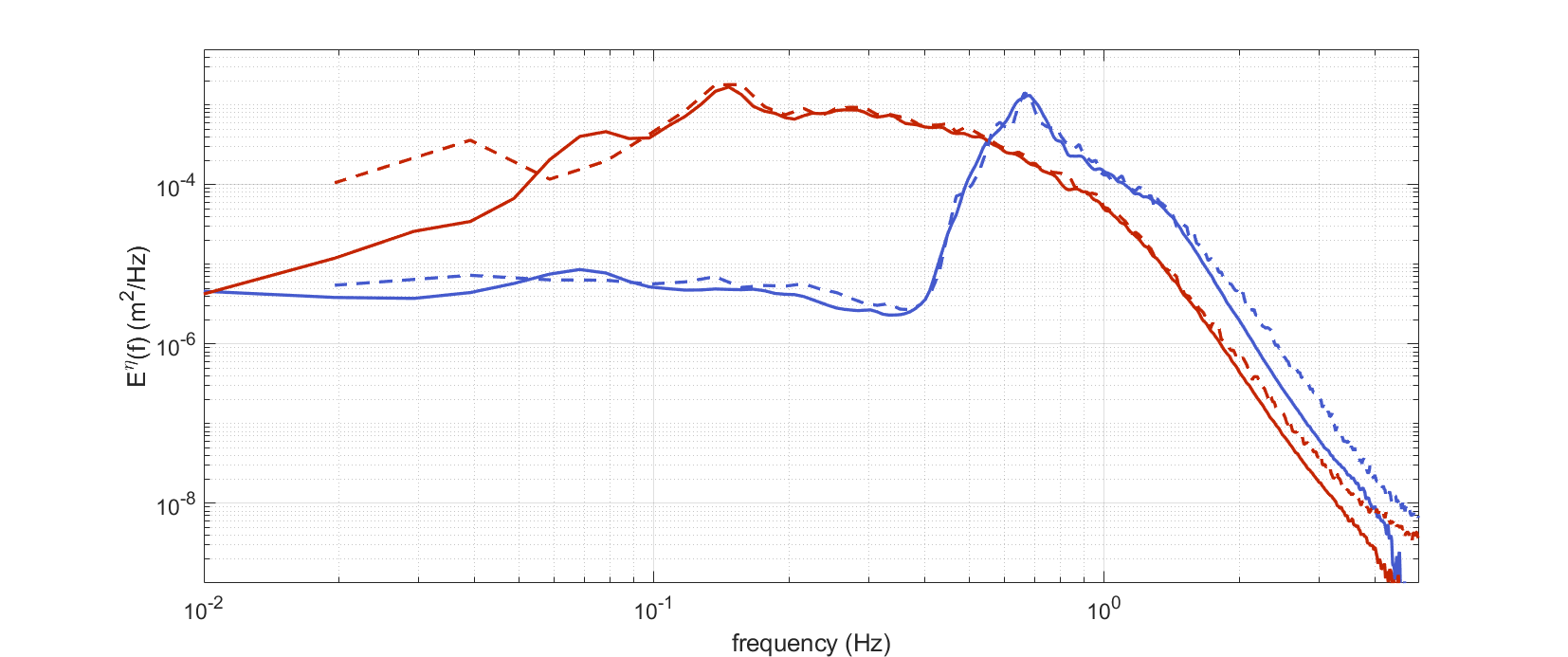}
(b)\includegraphics[width=\textwidth]{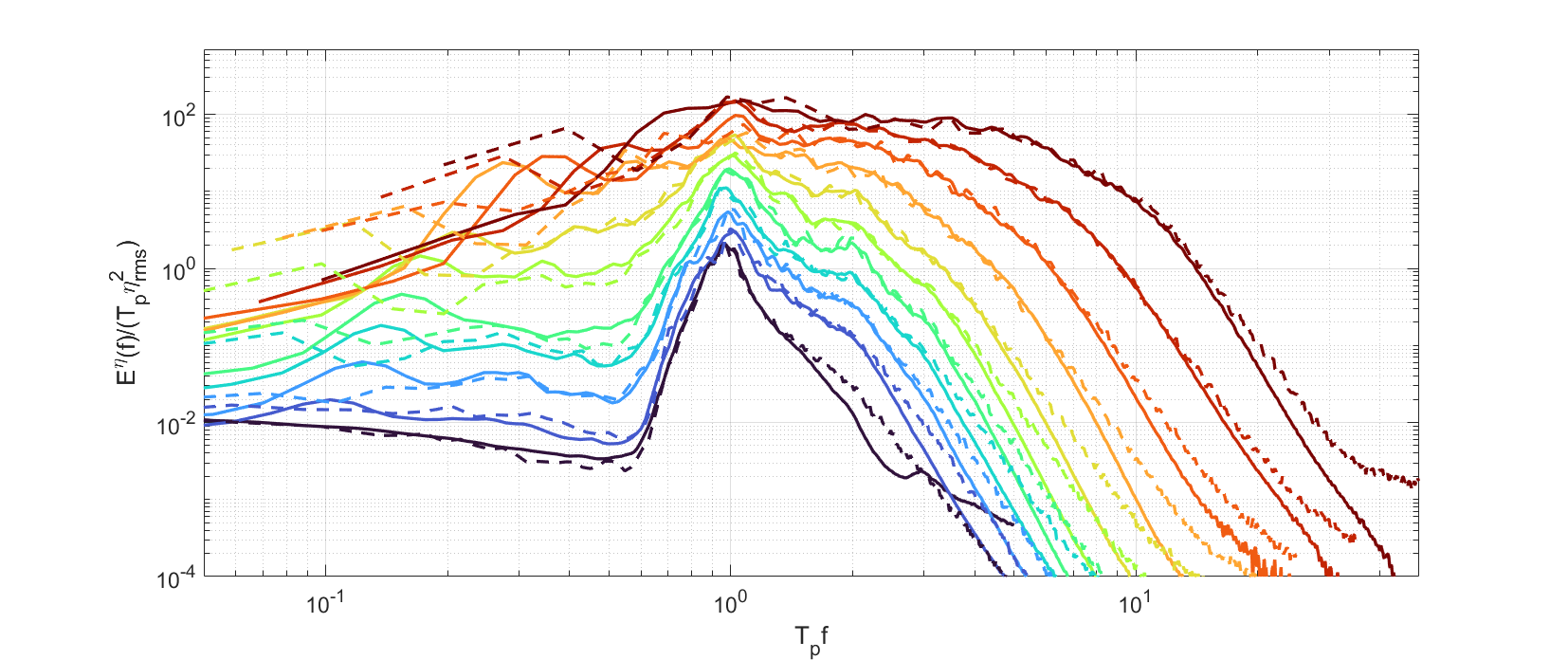}
\caption{\label{spf35comp}Frequency spectra for $H_{m0}=3.5$~cm as a function of $T_p$, compared between the measurement by capacitive probes distant from any wall by more than 5~m (average of 8 probes, dashed lines) and the stereoscopic reconstruction (solid lines) for the same data as Fig.~\ref{spf35}. Colors from black to dark red correspond to increasing value of $T_p$ from 1~s to 10~s. (a) are dimensional spectra (only for $T_p=1.5$ and $7$~s), and (b) are dimensionless spectra.}
\end{figure*}

One can see in Fig.~\ref{spf35comp} that the measurements by the capacitive probes and that of the stereoscopic reconstruction agree quite well up to frequencies about $1.5$~Hz. At higher frequencies, the stereoscopic reconstruction displays a slightly lower energy that may be due to some spatial low pass filtering effect due to image correlation that translates into a temporal low pass filtering through the propagation of the waves. Differences are also visible for low frequency peaks. This is due to the positioning of the probes as compared to nodes and antinodes of the discrete modes of the tank.

\section{\label{app:Mach}Reflection of an oblique soliton on a vertical wall}

The wavemakers are located at a single side of the basin, a generated oblique soliton therefore diffracts and degenerates into a bent soliton  
\citep{ryskamp2021evolution,leduquePhD2024}. 
The interaction of an oblique soliton with a vertical wall is called a Mach reflection \citep{Miles77b}, characterized by $\kappa$ (\ref{eq:kappa}). For $\kappa<1$, it produces a `stem' that evolves into a solitary wave propagating perpendicularly to the wall, with increasing length and decreasing amplitude. 

\begin{figure}
\centering
{{\includegraphics[height=129mm,viewport=120 200 500 650,clip]{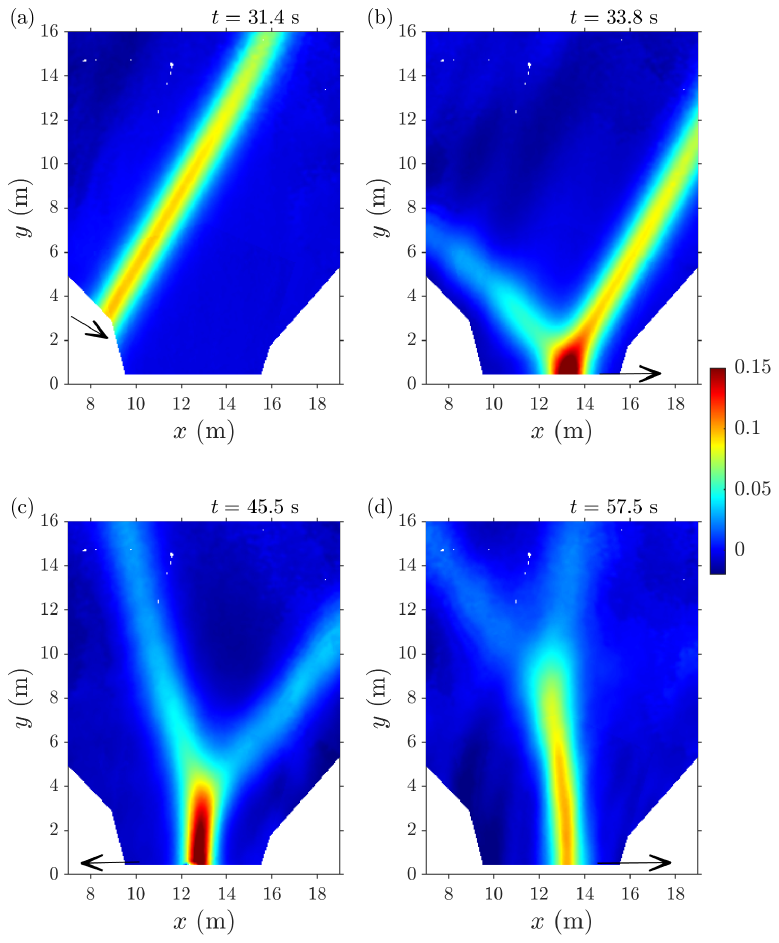}} }
\caption{
Snapshots of surface elevation ($\eta$ in m), recorded with the two sets of cameras, during Mach interaction:
a soliton of reduced amplitude $\epsilon=0.25$ and incidence $\theta_i=-30^\circ$ (a) produces a stem at the wall $y=0$ (b) that extends 
perpendicularly to the wall (c) along its propagation (d).
The black arrows indicate the main wave propagation direction. 
\label{MachInt2159}
}
\end{figure}

An example is shown in Fig.\,\ref{MachInt2159}.
A single soliton is generated with reduced amplitude $\epsilon=0.25$ and with $\kappa=0.86$ (Fig.\,\ref{MachInt2159}a). Its interaction with the wall $y=0$ produces a stem and a reflected wave  (Fig.\,\ref{MachInt2159}b).
At this stage the stem reduced amplitude is $\epsilon_w=0.57$ (about 2.3 times that of the initial soliton). 
The waves are then reflected by the end-wall at $x=27$\,m and recorded as they travel back towards the wavemakers (Fig.\,\ref{MachInt2159}c).
The stem is then about 4\,m long and $\epsilon_w=0.49$. 
It is shown after reflection on the wavemakers wall and travelling back into the video recording zone (Fig.\,\ref{MachInt2159}d).
The stem is then about 8\,m long with  $\epsilon_w=0.29$. 
Its incidence is close to $\theta=0$.

This underlines that, due to finite size effects, an oblique soliton can reduce into a solitary wave with $\theta\sim 0$.
More details on Mach interaction can be found in \cite{leduquePhD2024}.
\\

\bibliography{biblio.bib}

\begin{thebibliography}{57}%
\makeatletter
\providecommand \@ifxundefined [1]{%
 \@ifx{#1\undefined}
}%
\providecommand \@ifnum [1]{%
 \ifnum #1\expandafter \@firstoftwo
 \else \expandafter \@secondoftwo
 \fi
}%
\providecommand \@ifx [1]{%
 \ifx #1\expandafter \@firstoftwo
 \else \expandafter \@secondoftwo
 \fi
}%
\providecommand \natexlab [1]{#1}%
\providecommand \enquote  [1]{``#1''}%
\providecommand \bibnamefont  [1]{#1}%
\providecommand \bibfnamefont [1]{#1}%
\providecommand \citenamefont [1]{#1}%
\providecommand \href@noop [0]{\@secondoftwo}%
\providecommand \href [0]{\begingroup \@sanitize@url \@href}%
\providecommand \@href[1]{\@@startlink{#1}\@@href}%
\providecommand \@@href[1]{\endgroup#1\@@endlink}%
\providecommand \@sanitize@url [0]{\catcode `\\12\catcode `\$12\catcode
  `\&12\catcode `\#12\catcode `\^12\catcode `\_12\catcode `\%12\relax}%
\providecommand \@@startlink[1]{}%
\providecommand \@@endlink[0]{}%
\providecommand \url  [0]{\begingroup\@sanitize@url \@url }%
\providecommand \@url [1]{\endgroup\@href {#1}{\urlprefix }}%
\providecommand \urlprefix  [0]{URL }%
\providecommand \Eprint [0]{\href }%
\providecommand \doibase [0]{https://doi.org/}%
\providecommand \selectlanguage [0]{\@gobble}%
\providecommand \bibinfo  [0]{\@secondoftwo}%
\providecommand \bibfield  [0]{\@secondoftwo}%
\providecommand \translation [1]{[#1]}%
\providecommand \BibitemOpen [0]{}%
\providecommand \bibitemStop [0]{}%
\providecommand \bibitemNoStop [0]{.\EOS\space}%
\providecommand \EOS [0]{\spacefactor3000\relax}%
\providecommand \BibitemShut  [1]{\csname bibitem#1\endcsname}%
\let\auto@bib@innerbib\@empty
\bibitem [{\citenamefont {Scott~Russell}(1844)}]{scott1844report}%
  \BibitemOpen
  \bibfield  {author} {\bibinfo {author} {\bibfnamefont {J.}~\bibnamefont
  {Scott~Russell}},\ }\bibfield  {title} {\bibinfo {title} {Report on waves},\
  }in\ \href@noop {} {\emph {\bibinfo {booktitle} {Report of the fourteenth
  meeting of the British Association for the Advancement of Science}}}\
  (\bibinfo {year} {1844})\ pp.\ \bibinfo {pages} {311--390}\BibitemShut
  {NoStop}%
\bibitem [{\citenamefont {Gardner}\ \emph {et~al.}(1967)\citenamefont
  {Gardner}, \citenamefont {Greene}, \citenamefont {Kruskal},\ and\
  \citenamefont {Miura}}]{Gardner}%
  \BibitemOpen
  \bibfield  {author} {\bibinfo {author} {\bibfnamefont {C.}~\bibnamefont
  {Gardner}}, \bibinfo {author} {\bibfnamefont {J.}~\bibnamefont {Greene}},
  \bibinfo {author} {\bibfnamefont {M.}~\bibnamefont {Kruskal}},\ and\ \bibinfo
  {author} {\bibfnamefont {R.}~\bibnamefont {Miura}},\ }\bibfield  {title}
  {\bibinfo {title} {Method for solving the {Korteweg de Vries} equation},\
  }\href@noop {} {\bibfield  {journal} {\bibinfo  {journal} {Phys. Rev. Lett.}\
  }\textbf {\bibinfo {volume} {19}},\ \bibinfo {pages} {1095} (\bibinfo {year}
  {1967})}\BibitemShut {NoStop}%
\bibitem [{\citenamefont {Dauxois}\ and\ \citenamefont
  {Peyrard}(2006)}]{Dauxois}%
  \BibitemOpen
  \bibfield  {author} {\bibinfo {author} {\bibfnamefont {T.}~\bibnamefont
  {Dauxois}}\ and\ \bibinfo {author} {\bibfnamefont {M.}~\bibnamefont
  {Peyrard}},\ }\href@noop {} {\emph {\bibinfo {title} {Physics of Solitons}}}\
  (\bibinfo  {publisher} {Cambridge University Press},\ \bibinfo {address}
  {Cambridge},\ \bibinfo {year} {2006})\BibitemShut {NoStop}%
\bibitem [{\citenamefont {Gelash}\ \emph {et~al.}(2019)\citenamefont {Gelash},
  \citenamefont {Agafontsev}, \citenamefont {Zakharov}, \citenamefont {El},
  \citenamefont {Randoux},\ and\ \citenamefont {Suret}}]{gelash_bound_2019}%
  \BibitemOpen
  \bibfield  {author} {\bibinfo {author} {\bibfnamefont {A.}~\bibnamefont
  {Gelash}}, \bibinfo {author} {\bibfnamefont {D.}~\bibnamefont {Agafontsev}},
  \bibinfo {author} {\bibfnamefont {V.}~\bibnamefont {Zakharov}}, \bibinfo
  {author} {\bibfnamefont {G.}~\bibnamefont {El}}, \bibinfo {author}
  {\bibfnamefont {S.}~\bibnamefont {Randoux}},\ and\ \bibinfo {author}
  {\bibfnamefont {P.}~\bibnamefont {Suret}},\ }\bibfield  {title} {\bibinfo
  {title} {Bound {State} {Soliton} {Gas} {Dynamics} {Underlying} the
  {Spontaneous} {Modulational} {Instability}},\ }\href
  {https://doi.org/10.1103/PhysRevLett.123.234102} {\bibfield  {journal}
  {\bibinfo  {journal} {Physical Review Letters}\ }\textbf {\bibinfo {volume}
  {123}},\ \bibinfo {pages} {234102} (\bibinfo {year} {2019})}\BibitemShut
  {NoStop}%
\bibitem [{\citenamefont {Zakharov}(2009)}]{Zakharov}%
  \BibitemOpen
  \bibfield  {author} {\bibinfo {author} {\bibfnamefont {V.~E.}\ \bibnamefont
  {Zakharov}},\ }\bibfield  {title} {\bibinfo {title} {{Turbulence in
  integrable systems}},\ }\href@noop {} {\bibfield  {journal} {\bibinfo
  {journal} {Stud. Appl. Math.}\ }\textbf {\bibinfo {volume} {122}},\ \bibinfo
  {pages} {219} (\bibinfo {year} {2009})}\BibitemShut {NoStop}%
\bibitem [{\citenamefont {Zakharov}(1971)}]{zakharov1971kinetic}%
  \BibitemOpen
  \bibfield  {author} {\bibinfo {author} {\bibfnamefont {V.}~\bibnamefont
  {Zakharov}},\ }\bibfield  {title} {\bibinfo {title} {Kinetic equation for
  solitons},\ }\href@noop {} {\bibfield  {journal} {\bibinfo  {journal} {Sov.
  Phys. JETP}\ }\textbf {\bibinfo {volume} {33}},\ \bibinfo {pages} {538}
  (\bibinfo {year} {1971})}\BibitemShut {NoStop}%
\bibitem [{\citenamefont {El}\ and\ \citenamefont
  {Kamchatnov}(2005)}]{el2005kinetic}%
  \BibitemOpen
  \bibfield  {author} {\bibinfo {author} {\bibfnamefont {G.}~\bibnamefont
  {El}}\ and\ \bibinfo {author} {\bibfnamefont {A.}~\bibnamefont
  {Kamchatnov}},\ }\bibfield  {title} {\bibinfo {title} {Kinetic equation for a
  dense soliton gas},\ }\href@noop {} {\bibfield  {journal} {\bibinfo
  {journal} {Phys. Rev. Lett.}\ }\textbf {\bibinfo {volume} {95}},\ \bibinfo
  {pages} {204101} (\bibinfo {year} {2005})}\BibitemShut {NoStop}%
\bibitem [{\citenamefont {El}\ \emph {et~al.}(2011)\citenamefont {El},
  \citenamefont {Kamchatnov}, \citenamefont {Pavlov},\ and\ \citenamefont
  {Zykov}}]{el2011kinetic}%
  \BibitemOpen
  \bibfield  {author} {\bibinfo {author} {\bibfnamefont {G.~A.}\ \bibnamefont
  {El}}, \bibinfo {author} {\bibfnamefont {A.}~\bibnamefont {Kamchatnov}},
  \bibinfo {author} {\bibfnamefont {M.~V.}\ \bibnamefont {Pavlov}},\ and\
  \bibinfo {author} {\bibfnamefont {S.~A.}\ \bibnamefont {Zykov}},\ }\bibfield
  {title} {\bibinfo {title} {Kinetic equation for a soliton gas and its
  hydrodynamic reductions},\ }\href@noop {} {\bibfield  {journal} {\bibinfo
  {journal} {Journal of Nonlinear Science}\ }\textbf {\bibinfo {volume} {21}},\
  \bibinfo {pages} {151} (\bibinfo {year} {2011})}\BibitemShut {NoStop}%
\bibitem [{\citenamefont {El}(2021)}]{GEL}%
  \BibitemOpen
  \bibfield  {author} {\bibinfo {author} {\bibfnamefont {G.}~\bibnamefont
  {El}},\ }\bibfield  {title} {\bibinfo {title} {Soliton gas in integrable
  dispersive hydrodynamics},\ }\href@noop {} {\bibfield  {journal} {\bibinfo
  {journal} {J. Stat. Phys.: Theory Exp.}\ ,\ \bibinfo {pages} {114001}}
  (\bibinfo {year} {2021})}\BibitemShut {NoStop}%
\bibitem [{\citenamefont {Suret}\ \emph {et~al.}(2024)\citenamefont {Suret},
  \citenamefont {Randoux}, \citenamefont {Gelash}, \citenamefont {Agafontsev},
  \citenamefont {Doyon},\ and\ \citenamefont {El}}]{Suret}%
  \BibitemOpen
  \bibfield  {author} {\bibinfo {author} {\bibfnamefont {P.}~\bibnamefont
  {Suret}}, \bibinfo {author} {\bibfnamefont {S.}~\bibnamefont {Randoux}},
  \bibinfo {author} {\bibfnamefont {A.}~\bibnamefont {Gelash}}, \bibinfo
  {author} {\bibfnamefont {D.}~\bibnamefont {Agafontsev}}, \bibinfo {author}
  {\bibfnamefont {B.}~\bibnamefont {Doyon}},\ and\ \bibinfo {author}
  {\bibfnamefont {G.}~\bibnamefont {El}},\ }\bibfield  {title} {\bibinfo
  {title} {Soliton gas: Theory, numerics, and experiments},\ }\href@noop {}
  {\bibfield  {journal} {\bibinfo  {journal} {Phys.Rev. E}\ }\textbf {\bibinfo
  {volume} {109}},\ \bibinfo {pages} {061001} (\bibinfo {year}
  {2024})}\BibitemShut {NoStop}%
\bibitem [{\citenamefont {Zhang}\ and\ \citenamefont {Li}(2003)}]{Zhang}%
  \BibitemOpen
  \bibfield  {author} {\bibinfo {author} {\bibfnamefont {J.~E.}\ \bibnamefont
  {Zhang}}\ and\ \bibinfo {author} {\bibfnamefont {Y.}~\bibnamefont {Li}},\
  }\bibfield  {title} {\bibinfo {title} {{Bidirectional solitons on water}},\
  }\href@noop {} {\bibfield  {journal} {\bibinfo  {journal} {Phys. Rev. E}\
  }\textbf {\bibinfo {volume} {67}},\ \bibinfo {pages} {509} (\bibinfo {year}
  {2003})}\BibitemShut {NoStop}%
\bibitem [{\citenamefont {Kadomtsev}\ and\ \citenamefont
  {Petviashvili}(1970)}]{KP}%
  \BibitemOpen
  \bibfield  {author} {\bibinfo {author} {\bibfnamefont {B.}~\bibnamefont
  {Kadomtsev}}\ and\ \bibinfo {author} {\bibfnamefont {V.}~\bibnamefont
  {Petviashvili}},\ }\bibfield  {title} {\bibinfo {title} {On the stability of
  solitary waves in weakly dispersive media},\ }\href@noop {} {\bibfield
  {journal} {\bibinfo  {journal} {Sov. Phys.—Dokl.}\ }\textbf {\bibinfo
  {volume} {15}},\ \bibinfo {pages} {539} (\bibinfo {year} {1970})}\BibitemShut
  {NoStop}%
\bibitem [{\citenamefont {Kodama}(2010)}]{Kodama2010}%
  \BibitemOpen
  \bibfield  {author} {\bibinfo {author} {\bibfnamefont {Y.}~\bibnamefont
  {Kodama}},\ }\bibfield  {title} {\bibinfo {title} {{KP} solitons in shallow
  water},\ }\href@noop {} {\bibfield  {journal} {\bibinfo  {journal} {J. Phys.
  A: Math. Theor.}\ }\textbf {\bibinfo {volume} {43}},\ \bibinfo {pages}
  {434004} (\bibinfo {year} {2010})}\BibitemShut {NoStop}%
\bibitem [{\citenamefont {Ablowitz}\ and\ \citenamefont
  {Segur}(1981)}]{ablowitz1981solitons}%
  \BibitemOpen
  \bibfield  {author} {\bibinfo {author} {\bibfnamefont {M.~J.}\ \bibnamefont
  {Ablowitz}}\ and\ \bibinfo {author} {\bibfnamefont {H.}~\bibnamefont
  {Segur}},\ }\href@noop {} {\emph {\bibinfo {title} {Solitons and the inverse
  scattering transform}}}\ (\bibinfo  {publisher} {SIAM},\ \bibinfo {year}
  {1981})\BibitemShut {NoStop}%
\bibitem [{\citenamefont {Biondini}(2007)}]{biondini2007line}%
  \BibitemOpen
  \bibfield  {author} {\bibinfo {author} {\bibfnamefont {G.}~\bibnamefont
  {Biondini}},\ }\bibfield  {title} {\bibinfo {title} {Line soliton
  interactions of the {K}adomtsev-{P}etviashvili equation},\ }\href@noop {}
  {\bibfield  {journal} {\bibinfo  {journal} {Physical review letters}\
  }\textbf {\bibinfo {volume} {99}},\ \bibinfo {pages} {064103} (\bibinfo
  {year} {2007})}\BibitemShut {NoStop}%
\bibitem [{\citenamefont {Infeld}\ and\ \citenamefont
  {Rowlands}(2000)}]{infeld2000nonlinear}%
  \BibitemOpen
  \bibfield  {author} {\bibinfo {author} {\bibfnamefont {E.}~\bibnamefont
  {Infeld}}\ and\ \bibinfo {author} {\bibfnamefont {G.}~\bibnamefont
  {Rowlands}},\ }\href@noop {} {\emph {\bibinfo {title} {Nonlinear waves,
  solitons and chaos}}}\ (\bibinfo  {publisher} {Cambridge university press},\
  \bibinfo {year} {2000})\BibitemShut {NoStop}%
\bibitem [{\citenamefont {Kodama}(2018)}]{kodama2018solitons}%
  \BibitemOpen
  \bibfield  {author} {\bibinfo {author} {\bibfnamefont {Y.}~\bibnamefont
  {Kodama}},\ }\href@noop {} {\emph {\bibinfo {title} {Solitons in
  two-dimensional shallow water}}}\ (\bibinfo  {publisher} {SIAM},\ \bibinfo
  {year} {2018})\BibitemShut {NoStop}%
\bibitem [{\citenamefont {Miles}(1977{\natexlab{a}})}]{Miles77a}%
  \BibitemOpen
  \bibfield  {author} {\bibinfo {author} {\bibfnamefont {J.}~\bibnamefont
  {Miles}},\ }\bibfield  {title} {\bibinfo {title} {Obliquely interacting
  solitary waves},\ }\href@noop {} {\bibfield  {journal} {\bibinfo  {journal}
  {J. Fluid Mech.}\ }\textbf {\bibinfo {volume} {79}},\ \bibinfo {pages} {157}
  (\bibinfo {year} {1977}{\natexlab{a}})}\BibitemShut {NoStop}%
\bibitem [{\citenamefont {Miles}(1977{\natexlab{b}})}]{Miles77b}%
  \BibitemOpen
  \bibfield  {author} {\bibinfo {author} {\bibfnamefont {J.}~\bibnamefont
  {Miles}},\ }\bibfield  {title} {\bibinfo {title} {Resonantly interacting
  solitary waves},\ }\href@noop {} {\bibfield  {journal} {\bibinfo  {journal}
  {J. Fluid Mech.}\ }\textbf {\bibinfo {volume} {79}},\ \bibinfo {pages} {170}
  (\bibinfo {year} {1977}{\natexlab{b}})}\BibitemShut {NoStop}%
\bibitem [{\citenamefont {Wang}\ and\ \citenamefont {Pawlowicz}(2012)}]{WangP}%
  \BibitemOpen
  \bibfield  {author} {\bibinfo {author} {\bibfnamefont {C.}~\bibnamefont
  {Wang}}\ and\ \bibinfo {author} {\bibfnamefont {R.}~\bibnamefont
  {Pawlowicz}},\ }\bibfield  {title} {\bibinfo {title} {Oblique wave-wave
  interactions of nonlinear nearsurface internal waves in the strait of
  {Georgia}},\ }\href@noop {} {\bibfield  {journal} {\bibinfo  {journal} {J.
  Geophys. Res.: Oceans}\ }\textbf {\bibinfo {volume} {85}},\ \bibinfo {pages}
  {1548} (\bibinfo {year} {2012})}\BibitemShut {NoStop}%
\bibitem [{\citenamefont {Kodama}\ and\ \citenamefont {Yeh}(2016)}]{KodamaYeh}%
  \BibitemOpen
  \bibfield  {author} {\bibinfo {author} {\bibfnamefont {Y.}~\bibnamefont
  {Kodama}}\ and\ \bibinfo {author} {\bibfnamefont {H.}~\bibnamefont {Yeh}},\
  }\bibfield  {title} {\bibinfo {title} {The {KP} theory and {Mach}
  reflection},\ }\href@noop {} {\bibfield  {journal} {\bibinfo  {journal} {J.
  Fluid Mech.}\ }\textbf {\bibinfo {volume} {800}},\ \bibinfo {pages} {766}
  (\bibinfo {year} {2016})}\BibitemShut {NoStop}%
\bibitem [{\citenamefont {Bonnemain}\ \emph {et~al.}(2025)\citenamefont
  {Bonnemain}, \citenamefont {Biondini}, \citenamefont {Doyon}, \citenamefont
  {Roberti},\ and\ \citenamefont {El}}]{Bonnemain}%
  \BibitemOpen
  \bibfield  {author} {\bibinfo {author} {\bibfnamefont {T.}~\bibnamefont
  {Bonnemain}}, \bibinfo {author} {\bibfnamefont {G.}~\bibnamefont {Biondini}},
  \bibinfo {author} {\bibfnamefont {B.}~\bibnamefont {Doyon}}, \bibinfo
  {author} {\bibfnamefont {G.}~\bibnamefont {Roberti}},\ and\ \bibinfo {author}
  {\bibfnamefont {G.}~\bibnamefont {El}},\ }\bibfield  {title} {\bibinfo
  {title} {Two-dimensional stationary soliton gas},\ }\href@noop {} {\bibfield
  {journal} {\bibinfo  {journal} {Phys. Rev. Res.}\ }\textbf {\bibinfo {volume}
  {7}},\ \bibinfo {pages} {013143} (\bibinfo {year} {2025})}\BibitemShut
  {NoStop}%
\bibitem [{\citenamefont {Nazarenko}(2011)}]{nazarenko2011wave}%
  \BibitemOpen
  \bibfield  {author} {\bibinfo {author} {\bibfnamefont {S.}~\bibnamefont
  {Nazarenko}},\ }\href@noop {} {\emph {\bibinfo {title} {Wave turbulence}}},\
  Vol.\ \bibinfo {volume} {825}\ (\bibinfo  {publisher} {Springer Science \&
  Business Media},\ \bibinfo {year} {2011})\BibitemShut {NoStop}%
\bibitem [{\citenamefont {Newell}\ and\ \citenamefont
  {Rumpf}(2011)}]{newell_wave_2011}%
  \BibitemOpen
  \bibfield  {author} {\bibinfo {author} {\bibfnamefont {A.}~\bibnamefont
  {Newell}}\ and\ \bibinfo {author} {\bibfnamefont {B.}~\bibnamefont {Rumpf}},\
  }\bibfield  {title} {\bibinfo {title} {Wave {Turbulence}},\ }\href
  {https://doi.org/10.1146/annurev-fluid-122109-160807} {\bibfield  {journal}
  {\bibinfo  {journal} {Ann. Rev. Fluid Mech.}\ }\textbf {\bibinfo {volume}
  {43}},\ \bibinfo {pages} {59} (\bibinfo {year} {2011})}\BibitemShut {NoStop}%
\bibitem [{\citenamefont {Hasselmann}(1962)}]{Hasselmann}%
  \BibitemOpen
  \bibfield  {author} {\bibinfo {author} {\bibfnamefont {K.}~\bibnamefont
  {Hasselmann}},\ }\bibfield  {title} {\bibinfo {title} {On the non-linear
  energy transfer in gravity-wave spectrum. {P}art 1. {G}eneral theory},\
  }\href@noop {} {\bibfield  {journal} {\bibinfo  {journal} {J. Fluid Mech.}\
  }\textbf {\bibinfo {volume} {12}},\ \bibinfo {pages} {481} (\bibinfo {year}
  {1962})}\BibitemShut {NoStop}%
\bibitem [{\citenamefont {Hasselmann}\ \emph {et~al.}(1973)\citenamefont
  {Hasselmann}, \citenamefont {Barnett}, \citenamefont {Bouws}, \citenamefont
  {Carlson}, \citenamefont {Cartwright}, \citenamefont {Enke}, \citenamefont
  {Ewing}, \citenamefont {Gienapp}, \citenamefont {Hasselmann}, \citenamefont
  {Kruseman}, \citenamefont {Meerburg}, \citenamefont {M\"uller}, \citenamefont
  {Olbers}, \citenamefont {Richter}, \citenamefont {Sell},\ and\ \citenamefont
  {Walden}}]{jonswap}%
  \BibitemOpen
  \bibfield  {author} {\bibinfo {author} {\bibfnamefont {K.}~\bibnamefont
  {Hasselmann}}, \bibinfo {author} {\bibfnamefont {T.}~\bibnamefont {Barnett}},
  \bibinfo {author} {\bibfnamefont {E.}~\bibnamefont {Bouws}}, \bibinfo
  {author} {\bibfnamefont {H.}~\bibnamefont {Carlson}}, \bibinfo {author}
  {\bibfnamefont {D.}~\bibnamefont {Cartwright}}, \bibinfo {author}
  {\bibfnamefont {K.}~\bibnamefont {Enke}}, \bibinfo {author} {\bibfnamefont
  {J.}~\bibnamefont {Ewing}}, \bibinfo {author} {\bibfnamefont
  {H.}~\bibnamefont {Gienapp}}, \bibinfo {author} {\bibfnamefont
  {D.}~\bibnamefont {Hasselmann}}, \bibinfo {author} {\bibfnamefont
  {P.}~\bibnamefont {Kruseman}}, \bibinfo {author} {\bibfnamefont
  {A.}~\bibnamefont {Meerburg}}, \bibinfo {author} {\bibfnamefont
  {P.}~\bibnamefont {M\"uller}}, \bibinfo {author} {\bibfnamefont
  {D.}~\bibnamefont {Olbers}}, \bibinfo {author} {\bibfnamefont
  {K.}~\bibnamefont {Richter}}, \bibinfo {author} {\bibfnamefont
  {W.}~\bibnamefont {Sell}},\ and\ \bibinfo {author} {\bibfnamefont
  {H.}~\bibnamefont {Walden}},\ }\bibfield  {title} {\bibinfo {title}
  {{Measurements of wind-wave growth and swell decay during the Joint North Sea
  WAve Project (JONSWAP)}},\ }\href@noop {} {\bibfield  {journal} {\bibinfo
  {journal} {Ergaenzungsheft zur Deutschen Hydrographischen Zeitschrift, Reihe
  A}\ }\textbf {\bibinfo {volume} {28}},\ \bibinfo {pages} {1} (\bibinfo {year}
  {1973})}\BibitemShut {NoStop}%
\bibitem [{\citenamefont {Zakharov}\ \emph {et~al.}(2004)\citenamefont
  {Zakharov}, \citenamefont {Dias},\ and\ \citenamefont {Pushkarev}}]{Zak1D}%
  \BibitemOpen
  \bibfield  {author} {\bibinfo {author} {\bibfnamefont {V.}~\bibnamefont
  {Zakharov}}, \bibinfo {author} {\bibfnamefont {F.}~\bibnamefont {Dias}},\
  and\ \bibinfo {author} {\bibfnamefont {A.}~\bibnamefont {Pushkarev}},\
  }\bibfield  {title} {\bibinfo {title} {One-dimensional wave turbulence},\
  }\href@noop {} {\bibfield  {journal} {\bibinfo  {journal} {Phys. Rep.}\
  }\textbf {\bibinfo {volume} {398}},\ \bibinfo {pages} {1} (\bibinfo {year}
  {2004})}\BibitemShut {NoStop}%
\bibitem [{\citenamefont {Zakharov}\ and\ \citenamefont
  {Schulman}(1980)}]{zakharov1980degenerative}%
  \BibitemOpen
  \bibfield  {author} {\bibinfo {author} {\bibfnamefont {V.}~\bibnamefont
  {Zakharov}}\ and\ \bibinfo {author} {\bibfnamefont {E.}~\bibnamefont
  {Schulman}},\ }\bibfield  {title} {\bibinfo {title} {Degenerative dispersion
  laws, motion invariants and kinetic equations},\ }\href@noop {} {\bibfield
  {journal} {\bibinfo  {journal} {Physica D: Nonlinear Phenomena}\ }\textbf
  {\bibinfo {volume} {1}},\ \bibinfo {pages} {192} (\bibinfo {year}
  {1980})}\BibitemShut {NoStop}%
\bibitem [{\citenamefont {Zakharov}\ and\ \citenamefont
  {Schulman}(1988)}]{zakharov1988additional}%
  \BibitemOpen
  \bibfield  {author} {\bibinfo {author} {\bibfnamefont {V.}~\bibnamefont
  {Zakharov}}\ and\ \bibinfo {author} {\bibfnamefont {E.}~\bibnamefont
  {Schulman}},\ }\bibfield  {title} {\bibinfo {title} {On additional motion
  invariants of classical {H}amiltonian wave systems},\ }\href@noop {}
  {\bibfield  {journal} {\bibinfo  {journal} {Physica D: Nonlinear Phenomena}\
  }\textbf {\bibinfo {volume} {29}},\ \bibinfo {pages} {283} (\bibinfo {year}
  {1988})}\BibitemShut {NoStop}%
\bibitem [{\citenamefont {Janssen}\ and\ \citenamefont
  {Onorato}(2007)}]{janssen07}%
  \BibitemOpen
  \bibfield  {author} {\bibinfo {author} {\bibfnamefont {P.~A. E.~M.}\
  \bibnamefont {Janssen}}\ and\ \bibinfo {author} {\bibfnamefont
  {M.}~\bibnamefont {Onorato}},\ }\bibfield  {title} {\bibinfo {title} {The
  {Intermediate} {Water} {Depth} {Limit} of the {Zakharov} {Equation} and
  {Consequences} for {Wave} {Prediction}},\ }\href
  {https://doi.org/10.1175/JPO3128.1} {\bibfield  {journal} {\bibinfo
  {journal} {Journal of Physical Oceanography}\ }\textbf {\bibinfo {volume}
  {37}},\ \bibinfo {pages} {2389} (\bibinfo {year} {2007})}\BibitemShut
  {NoStop}%
\bibitem [{\citenamefont {Onorato}\ \emph {et~al.}(2009)\citenamefont
  {Onorato}, \citenamefont {Cavaleri}, \citenamefont {Fouques}, \citenamefont
  {Gramstad}, \citenamefont {Janssen}, \citenamefont {Monbaliu}, \citenamefont
  {Osborne}, \citenamefont {Pakozdi}, \citenamefont {Serio}, \citenamefont
  {Stansberg}, \citenamefont {Toffoli},\ and\ \citenamefont
  {Trulsen}}]{onorato09}%
  \BibitemOpen
  \bibfield  {author} {\bibinfo {author} {\bibfnamefont {M.}~\bibnamefont
  {Onorato}}, \bibinfo {author} {\bibfnamefont {L.}~\bibnamefont {Cavaleri}},
  \bibinfo {author} {\bibfnamefont {S.}~\bibnamefont {Fouques}}, \bibinfo
  {author} {\bibfnamefont {O.}~\bibnamefont {Gramstad}}, \bibinfo {author}
  {\bibfnamefont {P.~A. E.~M.}\ \bibnamefont {Janssen}}, \bibinfo {author}
  {\bibfnamefont {J.}~\bibnamefont {Monbaliu}}, \bibinfo {author}
  {\bibfnamefont {A.~R.}\ \bibnamefont {Osborne}}, \bibinfo {author}
  {\bibfnamefont {C.}~\bibnamefont {Pakozdi}}, \bibinfo {author} {\bibfnamefont
  {M.}~\bibnamefont {Serio}}, \bibinfo {author} {\bibfnamefont {C.~T.}\
  \bibnamefont {Stansberg}}, \bibinfo {author} {\bibfnamefont {A.}~\bibnamefont
  {Toffoli}},\ and\ \bibinfo {author} {\bibfnamefont {K.}~\bibnamefont
  {Trulsen}},\ }\bibfield  {title} {\bibinfo {title} {Statistical properties of
  mechanically generated surface gravity waves: a laboratory experiment in a
  three-dimensional wave basin},\ }\href
  {https://doi.org/10.1017/S002211200900603X} {\bibfield  {journal} {\bibinfo
  {journal} {Journal of Fluid Mechanics}\ }\textbf {\bibinfo {volume} {627}},\
  \bibinfo {pages} {235} (\bibinfo {year} {2009})}\BibitemShut {NoStop}%
\bibitem [{\citenamefont {Zve}\ \emph {et~al.}(2023)\citenamefont {Zve},
  \citenamefont {Swan},\ and\ \citenamefont {Hughes}}]{zve}%
  \BibitemOpen
  \bibfield  {author} {\bibinfo {author} {\bibfnamefont {E.}~\bibnamefont
  {Zve}}, \bibinfo {author} {\bibfnamefont {C.}~\bibnamefont {Swan}},\ and\
  \bibinfo {author} {\bibfnamefont {G.}~\bibnamefont {Hughes}},\ }\bibfield
  {title} {\bibinfo {title} {Crest-height statistics in finite water depth.
  {Part} 1: {The} role of the nonlinear interactions in uni-directional seas},\
  }\href {https://doi.org/10.1016/j.oceaneng.2023.116369} {\bibfield  {journal}
  {\bibinfo  {journal} {Ocean Engineering}\ }\textbf {\bibinfo {volume}
  {289}},\ \bibinfo {pages} {116369} (\bibinfo {year} {2023})}\BibitemShut
  {NoStop}%
\bibitem [{\citenamefont {Redor}\ \emph {et~al.}(2019)\citenamefont {Redor},
  \citenamefont {Barth\'elemy}, \citenamefont {Michallet}, \citenamefont
  {Onorato},\ and\ \citenamefont {Mordant}}]{RedorPRL}%
  \BibitemOpen
  \bibfield  {author} {\bibinfo {author} {\bibfnamefont {I.}~\bibnamefont
  {Redor}}, \bibinfo {author} {\bibfnamefont {E.}~\bibnamefont {Barth\'elemy}},
  \bibinfo {author} {\bibfnamefont {H.}~\bibnamefont {Michallet}}, \bibinfo
  {author} {\bibfnamefont {M.}~\bibnamefont {Onorato}},\ and\ \bibinfo {author}
  {\bibfnamefont {N.}~\bibnamefont {Mordant}},\ }\bibfield  {title} {\bibinfo
  {title} {Experimental evidence of a hydrodynamic soliton gas},\ }\href@noop
  {} {\bibfield  {journal} {\bibinfo  {journal} {Physical Review Letters}\
  }\textbf {\bibinfo {volume} {122}},\ \bibinfo {pages} {214502} (\bibinfo
  {year} {2019})}\BibitemShut {NoStop}%
\bibitem [{\citenamefont {Redor}\ \emph {et~al.}(2020)\citenamefont {Redor},
  \citenamefont {Barth\'elemy}, \citenamefont {Mordant},\ and\ \citenamefont
  {Michallet}}]{RedorEIF}%
  \BibitemOpen
  \bibfield  {author} {\bibinfo {author} {\bibfnamefont {I.}~\bibnamefont
  {Redor}}, \bibinfo {author} {\bibfnamefont {E.}~\bibnamefont {Barth\'elemy}},
  \bibinfo {author} {\bibfnamefont {N.}~\bibnamefont {Mordant}},\ and\ \bibinfo
  {author} {\bibfnamefont {H.}~\bibnamefont {Michallet}},\ }\bibfield  {title}
  {\bibinfo {title} {Analysis of soliton gas with large-scale video-based wave
  measurements},\ }\href@noop {} {\bibfield  {journal} {\bibinfo  {journal}
  {Exp. Fluids}\ }\textbf {\bibinfo {volume} {61}},\ \bibinfo {pages} {216}
  (\bibinfo {year} {2020})}\BibitemShut {NoStop}%
\bibitem [{\citenamefont {Redor}\ \emph {et~al.}(2021)\citenamefont {Redor},
  \citenamefont {Michallet}, \citenamefont {Mordant},\ and\ \citenamefont
  {Barth\'elemy}}]{RedorPRF}%
  \BibitemOpen
  \bibfield  {author} {\bibinfo {author} {\bibfnamefont {I.}~\bibnamefont
  {Redor}}, \bibinfo {author} {\bibfnamefont {H.}~\bibnamefont {Michallet}},
  \bibinfo {author} {\bibfnamefont {N.}~\bibnamefont {Mordant}},\ and\ \bibinfo
  {author} {\bibfnamefont {E.}~\bibnamefont {Barth\'elemy}},\ }\bibfield
  {title} {\bibinfo {title} {Experimental study of integrable turbulence in
  shallow water},\ }\href@noop {} {\bibfield  {journal} {\bibinfo  {journal}
  {Physical Review Fluids}\ }\textbf {\bibinfo {volume} {6}},\ \bibinfo {pages}
  {124801} (\bibinfo {year} {2021})}\BibitemShut {NoStop}%
\bibitem [{\citenamefont {Stokes}(1847)}]{stokes1847}%
  \BibitemOpen
  \bibfield  {author} {\bibinfo {author} {\bibfnamefont {G.}~\bibnamefont
  {Stokes}},\ }\bibfield  {title} {\bibinfo {title} {On the theory of
  oscillatory waves},\ }\href@noop {} {\bibfield  {journal} {\bibinfo
  {journal} {Trans. Cam. Philos. Soc.}\ }\textbf {\bibinfo {volume} {8}},\
  \bibinfo {pages} {441} (\bibinfo {year} {1847})}\BibitemShut {NoStop}%
\bibitem [{\citenamefont {Whitham}(1974)}]{whitham74}%
  \BibitemOpen
  \bibfield  {author} {\bibinfo {author} {\bibfnamefont {G.}~\bibnamefont
  {Whitham}},\ }\href@noop {} {\emph {\bibinfo {title} {Linear and nonlinear
  waves}}}\ (\bibinfo  {publisher} {John Wiley \& Sons},\ \bibinfo {year}
  {1974})\BibitemShut {NoStop}%
\bibitem [{\citenamefont {Dean}\ and\ \citenamefont
  {Dalrymple}(1991)}]{dean1991}%
  \BibitemOpen
  \bibfield  {author} {\bibinfo {author} {\bibfnamefont {R.}~\bibnamefont
  {Dean}}\ and\ \bibinfo {author} {\bibfnamefont {R.}~\bibnamefont
  {Dalrymple}},\ }\href@noop {} {\emph {\bibinfo {title} {Water wave mechanics
  for engineers and scientists}}},\ Vol.~\bibinfo {volume} {2}\ (\bibinfo
  {publisher} {world scientific publishing company},\ \bibinfo {year}
  {1991})\BibitemShut {NoStop}%
\bibitem [{\citenamefont {Toffoli}\ \emph {et~al.}(2007)\citenamefont
  {Toffoli}, \citenamefont {Monbaliu}, \citenamefont {Onorato}, \citenamefont
  {Osborne}, \citenamefont {Babanin},\ and\ \citenamefont
  {Bitner-Gregersen}}]{toffoli2007}%
  \BibitemOpen
  \bibfield  {author} {\bibinfo {author} {\bibfnamefont {A.}~\bibnamefont
  {Toffoli}}, \bibinfo {author} {\bibfnamefont {J.}~\bibnamefont {Monbaliu}},
  \bibinfo {author} {\bibfnamefont {M.}~\bibnamefont {Onorato}}, \bibinfo
  {author} {\bibfnamefont {A.~R.}\ \bibnamefont {Osborne}}, \bibinfo {author}
  {\bibfnamefont {A.~V.}\ \bibnamefont {Babanin}},\ and\ \bibinfo {author}
  {\bibfnamefont {E.}~\bibnamefont {Bitner-Gregersen}},\ }\bibfield  {title}
  {\bibinfo {title} {Second-{Order} {Theory} and {Setup} in {Surface} {Gravity}
  {Waves}: {A} {Comparison} with {Experimental} {Data}},\ }\href
  {https://doi.org/10.1175/2007JPO3634.1} {\bibfield  {journal} {\bibinfo
  {journal} {Journal of Physical Oceanography}\ }\textbf {\bibinfo {volume}
  {37}},\ \bibinfo {pages} {2726} (\bibinfo {year} {2007})}\BibitemShut
  {NoStop}%
\bibitem [{\citenamefont {Ursell}(1953)}]{ursell1953}%
  \BibitemOpen
  \bibfield  {author} {\bibinfo {author} {\bibfnamefont {F.}~\bibnamefont
  {Ursell}},\ }\bibfield  {title} {\bibinfo {title} {The long-wave paradox in
  the theory of gravity waves},\ }\href
  {https://doi.org/10.1017/S0305004100028887} {\bibfield  {journal} {\bibinfo
  {journal} {Mathematical Proceedings of the Cambridge Philosophical Society}\
  }\textbf {\bibinfo {volume} {49}},\ \bibinfo {pages} {685} (\bibinfo {year}
  {1953})}\BibitemShut {NoStop}%
\bibitem [{\citenamefont {Zhao}\ \emph {et~al.}(2024)\citenamefont {Zhao},
  \citenamefont {Wang},\ and\ \citenamefont {Liu}}]{zhao2024}%
  \BibitemOpen
  \bibfield  {author} {\bibinfo {author} {\bibfnamefont {K.}~\bibnamefont
  {Zhao}}, \bibinfo {author} {\bibfnamefont {Y.}~\bibnamefont {Wang}},\ and\
  \bibinfo {author} {\bibfnamefont {P.~L.-F.}\ \bibnamefont {Liu}},\ }\bibfield
   {title} {\bibinfo {title} {A guide for selecting periodic water wave
  theories - {Le} {Méhauté} (1976)’s graph revisited},\ }\href
  {https://doi.org/10.1016/j.coastaleng.2023.104432} {\bibfield  {journal}
  {\bibinfo  {journal} {Coastal Engineering}\ }\textbf {\bibinfo {volume}
  {188}},\ \bibinfo {pages} {104432} (\bibinfo {year} {2024})}\BibitemShut
  {NoStop}%
\bibitem [{\citenamefont {Leduque}\ \emph {et~al.}(2024)\citenamefont
  {Leduque}, \citenamefont {Barth\'elemy}, \citenamefont {Michallet},
  \citenamefont {Sommeria},\ and\ \citenamefont {Mordant}}]{leduqueEIF}%
  \BibitemOpen
  \bibfield  {author} {\bibinfo {author} {\bibfnamefont {T.}~\bibnamefont
  {Leduque}}, \bibinfo {author} {\bibfnamefont {E.}~\bibnamefont
  {Barth\'elemy}}, \bibinfo {author} {\bibfnamefont {H.}~\bibnamefont
  {Michallet}}, \bibinfo {author} {\bibfnamefont {J.}~\bibnamefont
  {Sommeria}},\ and\ \bibinfo {author} {\bibfnamefont {N.}~\bibnamefont
  {Mordant}},\ }\bibfield  {title} {\bibinfo {title} {Space–time statistics
  of {2D} soliton gas in shallow water studied by stereoscopic surface
  mapping},\ }\href@noop {} {\bibfield  {journal} {\bibinfo  {journal} {Exp.
  Fluids}\ }\textbf {\bibinfo {volume} {65}},\ \bibinfo {pages} {84} (\bibinfo
  {year} {2024})}\BibitemShut {NoStop}%
\bibitem [{\citenamefont {Leduque}(2024)}]{leduquePhD2024}%
  \BibitemOpen
  \bibfield  {author} {\bibinfo {author} {\bibfnamefont {T.}~\bibnamefont
  {Leduque}},\ }\emph {\bibinfo {title} {Étude expérimentale et numérique
  d'ondes non-linéaires multidirectionnelles en eau peu profonde: application
  aux gaz de solitons}},\ \href@noop {} {Ph.D. thesis},\ \bibinfo  {school}
  {Universit\'e Grenoble Alpes} (\bibinfo {year} {2024})\BibitemShut {NoStop}%
\bibitem [{\citenamefont {Campagne}\ \emph {et~al.}(2018)\citenamefont
  {Campagne}, \citenamefont {Hassaini}, \citenamefont {Redor}, \citenamefont
  {Sommeria}, \citenamefont {Valran}, \citenamefont {Viboud},\ and\
  \citenamefont {Mordant}}]{Campagne}%
  \BibitemOpen
  \bibfield  {author} {\bibinfo {author} {\bibfnamefont {A.}~\bibnamefont
  {Campagne}}, \bibinfo {author} {\bibfnamefont {R.}~\bibnamefont {Hassaini}},
  \bibinfo {author} {\bibfnamefont {I.}~\bibnamefont {Redor}}, \bibinfo
  {author} {\bibfnamefont {J.}~\bibnamefont {Sommeria}}, \bibinfo {author}
  {\bibfnamefont {T.}~\bibnamefont {Valran}}, \bibinfo {author} {\bibfnamefont
  {S.}~\bibnamefont {Viboud}},\ and\ \bibinfo {author} {\bibfnamefont
  {N.}~\bibnamefont {Mordant}},\ }\bibfield  {title} {\bibinfo {title} {Impact
  of dissipation on the energy spectrum of experimental turbulence of gravity
  surface waves},\ }\href@noop {} {\bibfield  {journal} {\bibinfo  {journal}
  {Physical Review Fluids}\ }\textbf {\bibinfo {volume} {3}},\ \bibinfo {pages}
  {044801} (\bibinfo {year} {2018})}\BibitemShut {NoStop}%
\bibitem [{\citenamefont {Deike}\ \emph {et~al.}(2015)\citenamefont {Deike},
  \citenamefont {Miquel}, \citenamefont {Gutierrez}, \citenamefont {Jamin},
  \citenamefont {Semin}, \citenamefont {Berhanu}, \citenamefont {Falcon},\ and\
  \citenamefont {Bonnefoy}}]{Deike}%
  \BibitemOpen
  \bibfield  {author} {\bibinfo {author} {\bibfnamefont {L.}~\bibnamefont
  {Deike}}, \bibinfo {author} {\bibfnamefont {B.}~\bibnamefont {Miquel}},
  \bibinfo {author} {\bibfnamefont {P.}~\bibnamefont {Gutierrez}}, \bibinfo
  {author} {\bibfnamefont {T.}~\bibnamefont {Jamin}}, \bibinfo {author}
  {\bibfnamefont {B.}~\bibnamefont {Semin}}, \bibinfo {author} {\bibfnamefont
  {M.}~\bibnamefont {Berhanu}}, \bibinfo {author} {\bibfnamefont
  {E.}~\bibnamefont {Falcon}},\ and\ \bibinfo {author} {\bibfnamefont
  {F.}~\bibnamefont {Bonnefoy}},\ }\bibfield  {title} {\bibinfo {title} {Role
  of the basin boundary conditions in gravity wave turbulence},\ }\href
  {https://doi.org/10.1017/jfm.2015.494} {\bibfield  {journal} {\bibinfo
  {journal} {J. Fluid Mech.}\ }\textbf {\bibinfo {volume} {781}},\ \bibinfo
  {pages} {196} (\bibinfo {year} {2015})}\BibitemShut {NoStop}%
\bibitem [{\citenamefont {Leckler}\ \emph {et~al.}(2015)\citenamefont
  {Leckler}, \citenamefont {Ardhuin}, \citenamefont {Peureux}, \citenamefont
  {Benetazzo}, \citenamefont {Bergamasco},\ and\ \citenamefont
  {Dulov}}]{Leckler}%
  \BibitemOpen
  \bibfield  {author} {\bibinfo {author} {\bibfnamefont {F.}~\bibnamefont
  {Leckler}}, \bibinfo {author} {\bibfnamefont {F.}~\bibnamefont {Ardhuin}},
  \bibinfo {author} {\bibfnamefont {C.}~\bibnamefont {Peureux}}, \bibinfo
  {author} {\bibfnamefont {A.}~\bibnamefont {Benetazzo}}, \bibinfo {author}
  {\bibfnamefont {F.}~\bibnamefont {Bergamasco}},\ and\ \bibinfo {author}
  {\bibfnamefont {V.}~\bibnamefont {Dulov}},\ }\bibfield  {title} {\bibinfo
  {title} {{Analysis and Interpretation of Frequency-Wavenumber Spectra of
  Young Wind Waves}},\ }\href {https://doi.org/10.1175/JPO-D-14-0237.1}
  {\bibfield  {journal} {\bibinfo  {journal} {J. Phys. Ocean.}\ }\textbf
  {\bibinfo {volume} {45}},\ \bibinfo {pages} {2484} (\bibinfo {year}
  {2015})}\BibitemShut {NoStop}%
\bibitem [{\citenamefont {Lenain}\ and\ \citenamefont
  {Melville}(2017)}]{Lenain}%
  \BibitemOpen
  \bibfield  {author} {\bibinfo {author} {\bibfnamefont {L.}~\bibnamefont
  {Lenain}}\ and\ \bibinfo {author} {\bibfnamefont {W.~K.}\ \bibnamefont
  {Melville}},\ }\bibfield  {title} {\bibinfo {title} {Measurements of the
  directional spectrum across the equilibrium saturation ranges of
  wind-generated surface waves.},\ }\href@noop {} {\bibfield  {journal}
  {\bibinfo  {journal} {Journal of Physical Oceanography}\ }\textbf {\bibinfo
  {volume} {47}},\ \bibinfo {pages} {2123} (\bibinfo {year}
  {2017})}\BibitemShut {NoStop}%
\bibitem [{\citenamefont {Zabusky}\ and\ \citenamefont {Kruskal}(1965)}]{ZK}%
  \BibitemOpen
  \bibfield  {author} {\bibinfo {author} {\bibfnamefont {N.}~\bibnamefont
  {Zabusky}}\ and\ \bibinfo {author} {\bibfnamefont {M.}~\bibnamefont
  {Kruskal}},\ }\bibfield  {title} {\bibinfo {title} {Interaction of "solitons"
  in a collisionless plasma and the recurrence of initial states},\ }\href@noop
  {} {\bibfield  {journal} {\bibinfo  {journal} {Physical Review Letters}\
  }\textbf {\bibinfo {volume} {15}},\ \bibinfo {pages} {240} (\bibinfo {year}
  {1965})}\BibitemShut {NoStop}%
\bibitem [{\citenamefont {Trillo}\ \emph {et~al.}(2016)\citenamefont {Trillo},
  \citenamefont {Deng}, \citenamefont {Biondini}, \citenamefont {Klein},
  \citenamefont {Clauss}, \citenamefont {Chabchoub},\ and\ \citenamefont
  {Onorato}}]{Trillo}%
  \BibitemOpen
  \bibfield  {author} {\bibinfo {author} {\bibfnamefont {S.}~\bibnamefont
  {Trillo}}, \bibinfo {author} {\bibfnamefont {G.}~\bibnamefont {Deng}},
  \bibinfo {author} {\bibfnamefont {G.}~\bibnamefont {Biondini}}, \bibinfo
  {author} {\bibfnamefont {M.}~\bibnamefont {Klein}}, \bibinfo {author}
  {\bibfnamefont {G.~F.}\ \bibnamefont {Clauss}}, \bibinfo {author}
  {\bibfnamefont {A.}~\bibnamefont {Chabchoub}},\ and\ \bibinfo {author}
  {\bibfnamefont {M.}~\bibnamefont {Onorato}},\ }\bibfield  {title} {\bibinfo
  {title} {{Experimental Observation and Theoretical Description of
  Multisoliton Fission in Shallow Water}},\ }\href@noop {} {\bibfield
  {journal} {\bibinfo  {journal} {Phys. Rev. Lett.}\ }\textbf {\bibinfo
  {volume} {117}},\ \bibinfo {pages} {144102} (\bibinfo {year}
  {2016})}\BibitemShut {NoStop}%
\bibitem [{\citenamefont {Tayfun}(1980)}]{Tayfun}%
  \BibitemOpen
  \bibfield  {author} {\bibinfo {author} {\bibfnamefont {M.~A.}\ \bibnamefont
  {Tayfun}},\ }\bibfield  {title} {\bibinfo {title} {{Narrow-band nonlinear sea
  waves}},\ }\href {https://doi.org/10.1029/JC085iC03p01548} {\bibfield
  {journal} {\bibinfo  {journal} {J. Geophys. Res. : Oceans}\ }\textbf
  {\bibinfo {volume} {85}},\ \bibinfo {pages} {1548} (\bibinfo {year}
  {1980})}\BibitemShut {NoStop}%
\bibitem [{\citenamefont {Socquet-Juglard}\ \emph {et~al.}(2005)\citenamefont
  {Socquet-Juglard}, \citenamefont {Dysthe}, \citenamefont {Trulsen},
  \citenamefont {Krogstad},\ and\ \citenamefont {Liu}}]{Socquet}%
  \BibitemOpen
  \bibfield  {author} {\bibinfo {author} {\bibfnamefont {H.}~\bibnamefont
  {Socquet-Juglard}}, \bibinfo {author} {\bibfnamefont {K.~B.}\ \bibnamefont
  {Dysthe}}, \bibinfo {author} {\bibfnamefont {K.}~\bibnamefont {Trulsen}},
  \bibinfo {author} {\bibfnamefont {H.~E.}\ \bibnamefont {Krogstad}},\ and\
  \bibinfo {author} {\bibfnamefont {J.}~\bibnamefont {Liu}},\ }\bibfield
  {title} {\bibinfo {title} {{Probability distributions of surface gravity
  waves during spectral changes}},\ }\href
  {https://doi.org/10.1017/S0022112005006312} {\bibfield  {journal} {\bibinfo
  {journal} {J. Fluid Mech.}\ }\textbf {\bibinfo {volume} {542}},\ \bibinfo
  {pages} {195} (\bibinfo {year} {2005})}\BibitemShut {NoStop}%
\bibitem [{\citenamefont {Zhang}\ \emph {et~al.}(2024)\citenamefont {Zhang},
  \citenamefont {Ma},\ and\ \citenamefont {Benoit}}]{zhang2024}%
  \BibitemOpen
  \bibfield  {author} {\bibinfo {author} {\bibfnamefont {J.}~\bibnamefont
  {Zhang}}, \bibinfo {author} {\bibfnamefont {Y.}~\bibnamefont {Ma}},\ and\
  \bibinfo {author} {\bibfnamefont {M.}~\bibnamefont {Benoit}},\ }\bibfield
  {title} {\bibinfo {title} {Statistical distributions of free surface
  elevation and wave height for out-of-equilibrium sea-states provoked by
  strong depth variations},\ }\href
  {https://doi.org/10.1016/j.oceaneng.2023.116645} {\bibfield  {journal}
  {\bibinfo  {journal} {Ocean Engineering}\ }\textbf {\bibinfo {volume}
  {293}},\ \bibinfo {pages} {116645} (\bibinfo {year} {2024})}\BibitemShut
  {NoStop}%
\bibitem [{\citenamefont {Tayfun}\ and\ \citenamefont
  {Alkhalidi}(2020)}]{Tayfun2020}%
  \BibitemOpen
  \bibfield  {author} {\bibinfo {author} {\bibfnamefont {M.}~\bibnamefont
  {Tayfun}}\ and\ \bibinfo {author} {\bibfnamefont {M.}~\bibnamefont
  {Alkhalidi}},\ }\bibfield  {title} {\bibinfo {title} {Distribution of
  sea-surface elevations in intermediate and shallow water depths.},\
  }\href@noop {} {\bibfield  {journal} {\bibinfo  {journal} {Coastal
  Engineering}\ }\textbf {\bibinfo {volume} {157}},\ \bibinfo {pages} {103651}
  (\bibinfo {year} {2020})}\BibitemShut {NoStop}%
\bibitem [{\citenamefont {{Le M\'ehaut\'e}}(1976)}]{lemehaute}%
  \BibitemOpen
  \bibfield  {author} {\bibinfo {author} {\bibfnamefont {B.}~\bibnamefont {{Le
  M\'ehaut\'e}}},\ }\href@noop {} {\emph {\bibinfo {title} {An Introduction to
  hydrodynamics and water waves}}}\ (\bibinfo  {publisher} {Springer Science \&
  Business Media},\ \bibinfo {year} {1976})\BibitemShut {NoStop}%
\bibitem [{\citenamefont {Jia}(2014)}]{Jia}%
  \BibitemOpen
  \bibfield  {author} {\bibinfo {author} {\bibfnamefont {Y.}~\bibnamefont
  {Jia}},\ }\emph {\bibinfo {title} {Numerical study of the KP solitons and
  higher order Miles theory of the Mach reflection in shallow water}},\
  \href@noop {} {\bibinfo {type} {Phd thesis}},\ \bibinfo  {school} {Ohio State
  University} (\bibinfo {year} {2014})\BibitemShut {NoStop}%
\bibitem [{\citenamefont {Goda}(1999)}]{goda}%
  \BibitemOpen
  \bibfield  {author} {\bibinfo {author} {\bibfnamefont {Y.}~\bibnamefont
  {Goda}},\ }\bibfield  {title} {\bibinfo {title} {A comparative review on the
  functional forms of directional wave spectrum},\ }\href@noop {} {\bibfield
  {journal} {\bibinfo  {journal} {Costal Eng. J.}\ }\textbf {\bibinfo {volume}
  {41}},\ \bibinfo {pages} {1} (\bibinfo {year} {1999})}\BibitemShut {NoStop}%
\bibitem [{\citenamefont {Ryskamp}\ \emph {et~al.}(2021)\citenamefont
  {Ryskamp}, \citenamefont {Maiden}, \citenamefont {Biondini},\ and\
  \citenamefont {Hoefer}}]{ryskamp2021evolution}%
  \BibitemOpen
  \bibfield  {author} {\bibinfo {author} {\bibfnamefont {S.}~\bibnamefont
  {Ryskamp}}, \bibinfo {author} {\bibfnamefont {M.~D.}\ \bibnamefont {Maiden}},
  \bibinfo {author} {\bibfnamefont {G.}~\bibnamefont {Biondini}},\ and\
  \bibinfo {author} {\bibfnamefont {M.~A.}\ \bibnamefont {Hoefer}},\ }\bibfield
   {title} {\bibinfo {title} {Evolution of truncated and bent gravity wave
  solitons: the {Mach} expansion problem},\ }\href@noop {} {\bibfield
  {journal} {\bibinfo  {journal} {J. Fluid Mech.}\ }\textbf {\bibinfo {volume}
  {909}},\ \bibinfo {pages} {A24} (\bibinfo {year} {2021})}\BibitemShut
  {NoStop}%
\end{thebibliography}%

\end{document}